\documentclass[aps,prd,preprint,superscriptaddress,showpacs,floatfix,nobibnotes]{revtex4-1}

\usepackage{latexsym}
\usepackage{amsmath}
\usepackage{amssymb}
\usepackage{graphicx}

\usepackage{bm}
\usepackage{epsfig}
\usepackage{subfigure}
\newcommand{\bea}{\begin{eqnarray}}
\newcommand{\eea}{\end{eqnarray}}

\newcommand{\be}{\begin{equation}}
\newcommand{\ee}{\end{equation}}
\newcommand{\nn}{\nonumber}

\begin{document}

\title{Momentum broadening in unstable quark-gluon plasma}

\author{M.E. Carrington}
\email[]{carrington@brandonu.ca} \affiliation{Department of Physics, Brandon University, Brandon, Manitoba, R7A 6A9 Canada}\affiliation{Winnipeg Institute for Theoretical Physics, Winnipeg, Manitoba}

\author{St. Mr\'owczy\'nski}
\email[]{Stanislaw.Mrowczynski@ncbj.gov.pl } \affiliation{Institute of Physics, Jan Kochanowski University, Kielce, Poland}\affiliation{National Centre for Nuclear Research, Warsaw, Poland}

\author{B. Schenke}
\email[]{bschenke@quark.phy.bnl.gov } \affiliation{Physics Department, Brookhaven National Laboratory, Upton, NY 11973, USA}

\date{October 14, 2016}

\begin{abstract}

Quark-gluon plasma produced at the early stage of ultrarelativistic heavy ion collisions is unstable, if weakly coupled, due to the anisotropy of its momentum distribution. Chromomagnetic fields are spontaneously generated and can reach magnitudes much exceeding typical values of the fields in equilibrated plasma. We consider a high energy test parton traversing an unstable plasma that is populated with strong fields. We study the momentum broadening parameter $\hat q$ which determines the radiative energy loss of the test parton. We develop a formalism which gives $\hat q$ as the solution of an initial value problem, and we focus on extremely oblate plasmas which are physically relevant for relativistic heavy ion collisions. The parameter $\hat q$ is found to be strongly dependent on time. For short times it is of the order of the equilibrium value, but at later times $\hat q$ grows exponentially due to the interaction of the test parton with unstable modes and becomes much bigger than the value in equilibrium. The momentum broadening is also strongly directionally dependent and is largest when the test parton velocity is transverse to the beam axis. Consequences of our findings for the phenomenology of jet quenching in relativistic heavy ion collisions are briefly discussed.

\end{abstract}

\pacs{12.38.Mh, 25.75.−q}

%Quark-gluon plasma, 12.38.Mh 
%Relativistic heavy-ion collisions, 25.75.−q

\maketitle

\normalsize

%%%%%%%%%%%%%%%%%%%%%%%%%%%%%%%%%%%%%%%%%%%%%%%%%%%%%%%%
\section{Introduction}
\label{introduction-section}
%%%%%%%%%%%%%%%%%%%%%%%%%%%%%%%%%%%%%%%%%%%%%%%%%%%%%%%%

Jet quenching is observed in relativistic heavy ions collisions at the Relativistic Heavy Ion Collider (RHIC) and the Large Hadron Collider (LHC). The experimental status of the phenomenon is reviewed in {\it e.g.} the article \cite{Norbeck:2014loa} and the whole field is introduced in the monograph \cite{Rak:2013yta}. There is mounting evidence that  jet quenching is caused by the interaction of jet partons with deconfined color charges and therefore the phenomenon is treated as a signal that quark-gluon plasma (QGP) is produced at an early stage of relativistic heavy ion collisions, see {\it e.g.} the reviews \cite{Majumder:2010qh,Qin:2015srf}.  

The energy loss of an isolated high energy (test) parton traversing QGP plays a key role in a quantitative understanding of jet quenching and has been intensively studied over a long period of time, see {\it e.g.} the review \cite{Peigne:2008wu}. The QGP produced in relativistic heavy ion collisions equilibrates rapidly and spends most of its lifetime in a state of local equilibrium, and therefore energy loss is usually computed in a locally equilibrated plasma which evolves hydrodynamically \cite{Majumder:2010qh,Qin:2015srf}. We therefore begin with a discussion of the basic concepts and characteristic scales of the problem, using the language appropriate for a thermalized system. Most of the energy of equilibrium plasma is carried by particles with typical momenta $p$ of the order of the temperature $p \sim T$ (hard modes). The momentum of the test parton is usually taken to be much bigger than $T$. There are also gauge fields (soft modes) in the plasma with momenta $k$ of order $\sim gT$, where $g$ is the coupling constant and is assumed to be small, $g \ll 1$. These soft modes are highly occupied due to the Bose-Einstein distribution $n_{\rm BE} (k) \sim T/k \sim 1/g$, and can be treated as classical fields. At leading order the soft modes carry only a small fraction of the total plasma energy but, because of their high occupation numbers, they interact frequently with plasma particles and the test parton and therefore play an important dynamical role.

The energetic test parton interacts with both hard and soft modes. Its interaction with the hard plasma particles can take the form of elastic binary collisions, or radiative processes which are sub-leading. The interactions with the soft collective modes come from both soft scatterings and radiation which is mostly collinear with the test parton velocity. In the case of light quarks and gluons,  radiative energy loss is expected to give the dominant contribution. For heavy test quarks radiative energy loss is presumably less important, due to the effect of the dead cone \cite{Peigne:2008wu} in which the emission of gluons is suppressed. 
 
Although the equilibration process of the QGP formed in relativistic heavy ion collisions is fast, there is a brief early phase when the plasma is out of equilibrium and the momentum distribution of the plasma constituents is anisotropic. The early state of the plasma system is therefore unstable due to chromomagnetic modes (see {\it e.g.} the review \cite{Mrowczynski:2016etf}), and the test parton spends some short period of time in a medium where chromomagnetic fields grow exponentially. These fields interact strongly with the test parton because they have large amplitudes, or -- using the language of quantum mechanics -- because there are highly populated soft modes. The consequence is that during this brief pre-equilibrium phase the test parton can lose a significant fraction of the total energy that it will ultimately give up to the plasma. 

We have recently studied  collisional energy loss in weakly coupled unstable QGP \cite{Carrington:2015xca}. Since this is an initial value problem, the results depend (in fact quite strongly) on the choice of initial conditions. The test parton typically loses energy as it traverses the plasma, but depending on the way the initial conditions are chosen, it can also gain energy. This is a well known phenomenon in electromagnetic plasmas, see {\it e.g.} \cite{Leemans:2006dx}. The energy loss (or gain) of the test parton depends exponentially on time, because of the presence of unstable modes in the plasma. In addition to the time dependence, the energy change is also strongly direction dependent.

In this paper we discuss the momentum broadening parameter $\hat{q}$ which gives the average transverse momentum broadening per unit length caused by the random kicks the test parton receives as it passes through the plasma medium. The parameter $\hat{q}$ determines the radiative component of the energy loss  \cite{Baier:1996sk} and therefore, together with our previous result for collisional energy loss \cite{Carrington:2015xca}, provides a description of the soft part of the energy loss of an energetic parton moving through an unstable plasma.  

The parameter $\hat{q}$ was computed in \cite{Romatschke:2006bb,Baier:2008js} for the case of quark-gluon plasma with an anisotropic momentum distribution.  However,  the plasma was treated as a static system and the exponential growth of the unstable modes was not taken into account. The numerical simulations of Ref. \cite{Dumitru:2007rp} show instead that  $\hat{q}$ receives a sizable contribution from these unstable modes and grows in time. Such behavior was also suggested in \cite{Majumder:2009cf}. 

Following the Langevin formulation of the problem which was proposed in \cite{Majumder:2009cf}, we compute the parameter $\hat q$ for a parton traveling through QGP with an oblate momentum distribution, which is relevant for relativistic heavy ion collisions. We find that the parameter $\hat q$ indeed grows exponentially in time due to the unstable modes. The formalism that we develop can be applied generally to either a QED plasma of ultrarelativistic electrons and positrons or a QGP. In the first part of the paper we use language that is applicable to a QED plasma, and in Sec.~\ref{QGP-section} we discuss how to modify our expressions so that they apply to QCD plasma. 

Throughout the paper we use natural units where $\hbar=c=k_B=1$.

%%%%%%%%%%%%%%%%%%%%%%%%%%%%%%%%%%%%%%%%%%%%%%%%%%%%%%%%
\section{Formulation of the problem}
\label{problem-section}
%%%%%%%%%%%%%%%%%%%%%%%%%%%%%%%%%%%%%%%%%%%%%%%%%%%%%%%%

We consider a high energy test particle which moves across a plasma system. Its motion is described by the Newtonian equation 
\be
\label{EOM}
\frac{d \vec{p}(t)}{d t} = \vec{F}\big(t,\vec{r}(t)\big) \,.
\ee
We use $\vec{r}(t)$, $\vec{u}$ and $\vec{p}(t)$, respectively, to denote the particle's  trajectory, velocity and momentum;
$ F(t,\vec{r}) \equiv e\big( \vec{E} (t,\vec{r}) + \vec{u} \times \vec{B}(t,\vec{r}) \big)$ is the Lorentz force acting on the test particle and;  $\vec{E} (t,\vec{r})$ and $\vec{B}(t,\vec{r})$ are electric and magnetic fields in the plasma. We consider a high energy test parton, which means that changes of its momentum are expected to be much smaller than the momentum itself. The velocity of the test parton is therefore assumed to be equal to the speed of light, $\vec{u}^2=1$, and  $\vec{u}$ is also assumed to be a constant vector. The trajectory of the test parton is $\vec{r}(t) = \vec{r}(0) + \vec{u}t$, and the solution of Eq.~(\ref{EOM}) reads
\be
\label{p-soln}
\vec{p}(t) = \vec{p}(0) + \int_0^t dt' \vec{F}\big(t',\vec{r}(t')\big)\, .
\ee
 
Within the Langevin approach, which is valid on time scales that are long compared to the correlation times between the underlying microscopic forces in the medium, one considers the ensemble average $\langle p^i (t) \, p^j (t)\rangle$ which equals
\be
\label{cor-pi-pj}
\langle p^i (t) \, p^j (t)\rangle = 
\langle p^i(0)\, p^j(0)\rangle + \int_0^t dt_1 \int_0^t dt_2 \langle F^i (t_1,\vec{r}_1) \, F^j(t_2,\vec{r}_2) \rangle \,.
\ee
We assume here that the force $ \vec{F}(t,\vec{r}) $ is independent of the initial momentum $ \vec{p}(0)$ and that 
the ensemble average of the force vanishes, which means $\langle \vec{F} (t,\vec{r}) \rangle = 0$. The fields in the expression (\ref{cor-pi-pj}) are evaluated along the trajectory of the test parton and we use the notation $\vec r_i \equiv \vec r(t_i)$ with $i = 1,\,2$.

We are interested in the parameter $\hat{q}$ which measures the momentum broadening per unit time of a test parton in the direction transverse to its initial velocity. The parameter is defined as  
\be
\label{qhat1}
\hat{q}(t) \equiv  \frac{d}{dt} (\delta^{ij} - u^i u^i ) \langle p^i (t) \, p^j (t)\rangle .
\ee
Substituting the correlation function (\ref{cor-pi-pj}) into the definition (\ref{qhat1}) we obtain
\bea
\label{qhat3}
\hat q (t) = e^2 \frac{d}{dt} \int_0^t dt_1 \int_0^t dt_2\bigg[ \langle \vec E(t_1,\vec r_1 ) \cdot \vec E(t_2,\vec r_2) - \vec u \cdot \vec E(t_1,\vec r_1) \,\vec u \cdot \vec E(t_2,\vec r_2) \rangle 
\nonumber\\
\langle \vec B(t_1,\vec r_1) \cdot \vec B(t_2,\vec r_2) - \vec u \cdot \vec B(t_1,\vec r_1) \,\vec u \cdot \vec B(t_2,\vec r_2) \rangle\nonumber\\[2mm]
-\langle \vec u \cdot \big[\vec E(t_1,\vec r_1) \times \vec B(t_2,\vec r_2)\big] \rangle 
+ \langle \vec u \cdot \big[\vec B(t_1,\vec r_1) \times \vec E(t_2,\vec r_2)\big] \rangle
 \bigg]  .
\eea
Thus we find that the parameter  $\hat{q}$ is determined by a set of field correlation functions, which are calculated in the next two sections.

When momentum broadening results from multiple independent collisions of the test parton with plasma constituents, the parameter $\hat{q}$ is time-independent and the total transverse momentum broadening equals $\langle p^2_T \rangle_{\rm tot} = \hat{q} \,L$ where $L$ is the path length of the test parton in the plasma, which is assumed to be static. In our approach the test parton interacts with a time-dependent chromodynamic field generated in the plasma. The parameter $\hat{q}$ is therefore time dependent and the momentum broadening equals
\be
\label{qhat-ess}
\langle p^2_T(t) \rangle = \int_0^t dt'  \hat{q}(t') \,.
\ee
In the equilibrium limit, the parameter $\hat{q}$ is time-independent, as discussed in detail in Appendix~\ref{equib-appendix}. Since the upper limit of the integral in (\ref{qhat-ess}) is proportional to $L$ for a relativistic parton, we find that in equilibrium the total momentum broadening is proportional to $L$. 
However, such behavior is rather exceptional. In the case of unstable plasmas, which we are primarily interested in, the momentum broadening $\langle p^2_T(t) \rangle$ can grow exponentially with $L$ if the exponentially growing modes are mostly responsible for the momentum broadening. This is in fact the main result of our study.

%%%%%%%%%%%%%%%%%%%%%%%%%%%%%%%%%%%%%%%%%%%%%%%%%%%%%%%%
\section{Fields in the plasma}
\label{correlator-section}
%%%%%%%%%%%%%%%%%%%%%%%%%%%%%%%%%%%%%%%%%%%%%%%%%%%%%%%%

Our aim in this section is to derive expressions for the electric and magnetic fields present in the plasma which enter the field correlators in Eq.~(\ref{qhat3}). We start with a consideration of the situation before the test parton arrives. We have a non-equilibrated plasma populated by fields (soft modes) that are generated by plasma constituents (hard modes) which are described by the phase-space distribution function $n_\sigma(t,\vec r,\vec p)$. This function obeys the Vlasov equation
\bea
\label{vlasov-1}
\bigg[\frac{\partial }{\partial t} + \vec v \cdot \vec \nabla 
+q_\sigma \Big(\vec E (t,\vec r)+\vec v \times \vec B (t,\vec r) \Big) \cdot \vec\nabla_p\bigg] n_\sigma(t,\vec r,\vec p) = 0 \,,
\eea
where $q_\sigma = \pm e$ is the charge of plasma constituents (electrons and positrons). These particles  are assumed to be massless and thus the velocity related to the momentum $\vec p$ is $\vec v = \vec p/p$ with $p \equiv |\vec p|$. 

We expand the distribution function $n_\sigma(t,\vec r,\vec p)$ around a stationary, homogeneous and charge neutral but anisotropic state whose distribution function is denoted $n_\sigma(\vec p)$. The distribution function is therefore written as
\bea
\label{expansion}
n_\sigma(t,\vec r,\vec p) = n_\sigma(\vec p)+\delta n_\sigma(t,\vec r,\vec p)\, ,
\eea
where $|\delta n_\sigma(t,\vec r,\vec p)| \ll n_\sigma(\vec p)$ and $|\nabla_p^i \delta n_\sigma(t,\vec r,\vec p)| \ll |\nabla_p^i n_\sigma(\vec p)|$. The fields $\vec E$ and $\vec B$ are considered first order in the expansion (\ref{expansion}) because, due to charge neutrality, they would be zero everywhere in the homogeneous system. Equation~(\ref{vlasov-1}) expanded up to the first order is
\bea
\label{vlasov-2}
\bigg(\frac{\partial }{\partial t} + \vec v \cdot \vec \nabla\bigg)\delta n_\sigma(t,\vec r,\vec p) 
+q_\sigma 
 \Big(\vec E (t,\vec r)+\vec v \times \vec B (t,\vec r) \Big) \cdot \vec \nabla_p n_\sigma(\vec p) = 0 \,.
\eea

The fields in the plasma are self-consistently generated by the moving particles according to Maxwell's equations:
\bea
\label{max-coord-homo}
\vec\nabla \cdot \vec B=0 \,,~~~~~
\vec\nabla \times \vec E + \frac{\partial \vec B}{\partial t}=0\, ,
\\
\label{max-coord-inhomo}
\vec\nabla \cdot \vec E = \rho \,,~~~~~
\vec\nabla \times \vec B - \frac{\partial \vec E}{\partial t} = \vec j \,,
\eea
where the charge density $\rho$ and current density $\vec j$ are given as
\bea
\label{ind-density}
\rho(t,\vec r) &=& \sum _\sigma q_\sigma \int d^3 p \,\delta n_\sigma(t,\vec r,\vec p) \,,
\\
\label{ind-current}
\vec j(t,\vec r) &=& \sum _\sigma q_\sigma \int d^3 p \, \vec v \, \delta n_\sigma(t,\vec r,\vec p) \,.
\eea
We use Heaviside-Lorentz electromagnetic units which are usually used in quantum field theory. 

The coupled set of equations (\ref{vlasov-2} - \ref{max-coord-inhomo}) can be solved self consistently. The physical interpretation is as follows. In the Maxwell equations (\ref{max-coord-homo}, \ref{max-coord-inhomo}) the current is viewed as the source for the fields, and in the Vlasov equation (\ref{vlasov-2}) the fields exert a force on the moving charges that make up the current.  

There is a subtle point associated with this procedure. When a parton enters an unequilibrated plasma the total source is the sum of the contribution produced by local deviations in the distribution of plasma particles from a stationary homogeneous state (see Eqs.~(\ref{ind-density}, \ref{ind-current})), and the contribution from the parton itself. The electric and magnetic fields determined from Maxwell's equations are therefore combinations of the induced field, with which the parton interacts, and the parton's own field. 
The interaction of the parton with its own field should not contribute to the momentum broadening coefficient. Equivalently, the momentum broadening coefficient should be zero for a parton moving through vacuum. In section \ref{integrand-section} we verify that this condition is satisfied. 

The first step in solving equations (\ref{vlasov-2} - \ref{max-coord-inhomo}) is to remove the differential operators by Fourier transforming.  The ordinary (two sided) Fourier transform converts a function of time to a function of frequency using an integral over time which extends from minus infinity to infinity. This is not what we want to do. We anticipate the fact that the parton will enter our system at $t_0=0$. We want to consider the interaction of this parton with the fields of the plasma. The unequilibrated plasma is not time translation invariant and we therefore need to develop a theoretical formalism in which the evolution of the parton can be calculated in a way that depends on the initial conditions.  In order to do this, we perform a one sided Fourier transform, which uses a time integral from zero to infinity. For a generic function $h$ the one sided Fourier transform and its inverse are defined as
\bea
\label{1side-forward}
h(\omega,\vec k) &=& \int_0^\infty dt \int d^3r \,
e^{i(\omega t -  \vec k\cdot \vec r)} h(t,\vec r) ,
\\[2mm]
\label{1side-inverse}
h(t, \vec r) &=& \int_{-\infty +i\sigma}^{\infty +i\sigma}
\frac{d\omega}{2\pi} \int\frac{d^3k}{(2\pi)^3} \,
e^{-i(\omega t - \vec k\cdot \vec r)} h(\omega,\vec k) .
\eea
The inverse transformation (\ref{1side-inverse}) involves the real parameter $\sigma > 0$ which is chosen so that the integral over $\omega$ is taken along a straight line in the complex $\omega$-plane, parallel to the real axis and above all singularities of $h(\omega,\vec k)$. Integrals over $\vec r$ and $\vec k$ are always taken over full $\vec r-$ and $\vec k-$space. 
Using equations (\ref{1side-forward}, \ref{1side-inverse}) we will take the one sided Fourier transform of our self-consistent set of equations, solve for the fields as functions of frequency and 3-momentum, and then perform the reverse transform to obtain the fields as functions of time and position. These expressions will be restricted to positive times, but this is exactly what we need in equation (\ref{qhat3}) which determines the momentum broadening coefficient. 

After taking the one sided Fourier transform, the Maxwell equations (\ref{max-coord-homo}, \ref{max-coord-inhomo}) become
\bea
\label{max-k-1}
&& i \vec k \cdot \vec B (\omega,\vec k) = 0 \,,
\\
\label{max-k-2}
&& i \vec k \times \vec E(\omega,\vec k) = i\omega \vec B(\omega,\vec k) + \vec B_{0}(\vec k) \,,
\\
\label{max-k-3}
&& i \vec k  \cdot \vec E(\omega, \vec k) = \rho (\omega,\vec k)\,,
\\
\label{max-k-4}
&& i \vec k \times \vec B (\omega,\vec k) = \vec j(\omega,\vec k) -i \omega \vec E(\omega,\vec k) - \vec E_{0}(\vec k) \,,
\eea
and the Vlasov equation (\ref{vlasov-2}) has the form
\bea
\label{vlasov-3}
-i (\omega - \vec k\cdot\vec v)\delta n_\sigma(\omega,\vec k,\vec p) 
+ q_\sigma \Big(\vec E (\omega,\vec k)+\vec v\times\vec B (\omega,\vec k) \Big)\cdot\nabla_p n_\sigma(\vec p) 
= \delta n_{0\sigma}(\vec k,\vec p)\,,
\eea 
where we have defined
\bea
\vec E_0(\vec r) \equiv \vec E(t=0,\vec r)\,, ~~~~
\vec B_0(\vec r) \equiv \vec B(t=0,\vec r)\,, ~~~~
\delta n_{\sigma 0}(\vec r,\vec p) \equiv \delta n_\sigma(t=0,\vec r,\vec p) \,.
\eea
Equations (\ref{max-k-1} - \ref{vlasov-3}) depend on the initial conditions because the one sided Fourier transform was used. Formally they reduce to the usual expressions for a time translation invariant system when the initial conditions are set to zero.

We can rewrite the Maxwell equations (\ref{max-k-2}, \ref{max-k-4}) in the form
\bea
\label{max-1s}
\big(\Delta_{\rm{bare}}^{-1}\big)^{ij}(\omega,\vec k) \, E^j(\omega,\vec k) = -i \omega j^i(\omega,\vec k) 
+ i \omega E_0^i(\vec k) - i \big(\vec k\times \vec B_0(\vec k) \big)^i
\eea
where the matrix $\big(\Delta_{\rm{bare}}^{-1}\big)^{ij}(\omega,\vec k)$ equals 
\bea
\label{bare-prop-inv}
\big(\Delta_{\rm{bare}}^{-1}\big)^{ij} (\omega,\vec k) =(\omega^2-\vec k^2)  \delta^{ij} + k^i k^j \,,
\eea
and 
\bea
\label{bare-prop}
\Delta_{\rm{bare}}^{ij}(\omega,\vec k)=\frac{1}{\omega^2-\vec k^2}\bigg( \delta^{ij} - \frac{k^i k^j}{\omega^2}\bigg) \,.
\eea

In quantum field theory $\Delta_{\rm{bare}}^{ij}(\omega,\vec k)$ is  the bare gauge field (photon) propagator in temporal axial gauge ($A^0=0$).  We emphasize however that in the case of QED our entire derivation, and the resulting formula for the momentum broadening parameter, is clearly gauge independent. This is evident from the fact that the calculation is formulated in terms of the gauge invariant fields $\vec E$ and $\vec B$ together with the source variables $\rho$, $\vec j$, $n_\sigma$ and $\delta n_\sigma$ which are also gauge independent. In a QGP the situation is more subtle because the analogs of the quantities $\vec E$,  $\vec B$, $\rho$, $\vec j$ and $\delta n_\sigma$ are gauge dependent in QCD. This point is discussed in section \ref{QGP-section}, where we show that the results for $\hat q$ obtained from our formalism are also gauge invariant in QCD. 

To obtain the electric field $\vec E$ from Eq.~(\ref{max-1s}) one needs an expression for the source current which enters the equation on the right side. In order to produce a self-consistent expression for the field, we obtain this current by solving the Vlasov equation (\ref{vlasov-3}) for $\delta n_\sigma(\omega,\vec k,\vec p)$ and substituting the solution into the definition of the current (\ref{ind-current}).  Eliminating the magnetic field using Eq.~(\ref{max-k-2}), the current equals
\bea
\label{vlasov-1s}
-i \omega j^i(\omega,\vec k) &=& \Pi^{ij}(\omega,\vec k)\,E^j(\omega,\vec k)
\\[2mm]
&+& \omega \sum_\sigma q_\sigma \int \frac{d^3 p}{(2\pi)^3}\,\frac{\vec v}{\omega-\vec v\cdot\vec k+i\epsilon}\, \bigg(\delta n_{0\sigma}(\vec k,\vec p) - i \frac{q_\sigma}{2\omega}\big(\vec v \times \vec B_0(\vec k)\big)\cdot \vec\nabla_p n_\sigma(\vec p)\bigg) \,,
\nonumber
\eea
where 
\bea
\label{PI}
\Pi^{ij}(\omega,\vec k) = 
-\omega  \int {d^3p \over (2\pi)^3} \,\frac{v^i}{\omega -  \vec v \cdot \vec k+i\epsilon}
\bigg(\Big(1-\frac{\vec k\cdot \vec v}{\omega}\Big) \delta^{jl}
+ \frac{v^j k^l}{\omega} \bigg) \nabla_p^l \sum_\sigma q_\sigma^2 n_\sigma(\vec p) \,,
\eea
which is the polarization tensor of an anisotropic plasma. If the system were translationally invariant in time, one would use a two sided Fourier transformation instead of the one sided transform, which is formally equivalent to dropping the initial conditions in Eq.~(\ref{vlasov-1s}). In this case, we  recover the familiar expression $-i \omega j^i(\omega,\vec k) = \Pi^{ij}(\omega,\vec k)\,E^j(\omega,\vec k)$, which says that the polarization tensor connects the electric field with the induced current that produced it. 

Integrating Eq. (\ref{PI}) by parts one obtains an expression that is frequently more useful
\bea
\label{PI2}
\Pi^{ij}(\omega,\vec k) = \sum_\sigma q_\sigma^2 \int\frac{d^3p}{(2\pi)^3}\frac{n_\sigma(\vec p)}{p} \bigg[
\delta_{ij}+\frac{k^i v^j+v^i k^j}{\omega-\vec v\cdot\vec k +i\epsilon}-\frac{(\omega^2-k^2)v^iv^j}{(\omega-\vec v\cdot\vec k +i\epsilon)^2}\bigg]\,.
\eea
Since we are assuming ultrarelativistic (massless) plasma constituents, the integral over the magnitude $p$ and angular integrals factorizes. We take advantage of this to rewrite the formula (\ref{PI2}) as 
\bea
\label{PI3}
\Pi^{ij}(\omega,\vec k) = \frac{m^2}{2} \int \frac{d\Omega}{4\pi} \bigg[
\delta_{ij}+\frac{k^i v^j+v^i k^j}{\omega-\vec v\cdot\vec k +i\epsilon}-\frac{(\omega^2-k^2)v^iv^j}{(\omega-\vec v\cdot\vec k +i\epsilon)^2}\bigg]\,,
\eea
where we have defined
\bea
\label{mass-def-new}
m^2 \equiv 2 \sum_\sigma q_\sigma^2 \int\frac{d^3p}{(2\pi)^3}\frac{n_\sigma(\vec p)}{p}\,.
\eea
The parameter $m$ is a characteristic mass scale that we use to define our units (in numerical calculations we set $m=1$). Physically the scale $m$ is related to the Debye mass; this is discussed in Appendix \ref{mass-def-section}. We comment that although the polarization tensor is usually used when discussing the QCD plasma, for an electromagentic plasma it is common to use the dielectric tensor $\varepsilon^{ij}(\omega, \vec k)$ which is related to $\Pi^{i j}(\omega,\vec k)$ as $\varepsilon^{ij} (\omega,\vec k) = \delta^{ij} - \omega^{-2} \Pi^{i j}(\omega,\vec k) $. 

With the current (\ref{vlasov-1s}) substituted into Eq.~(\ref{max-1s}), we obtain an equation that contains only fields and the initial fluctuation of the distribution function
\bea
\label{elec1}
&&\big( \Delta^{-1}(\omega,\vec k) \big)^{ij}\,E^j(\omega,\vec k) 
= i \omega E_0^i(\vec k) - i(\vec k\times \vec B_0(\vec k))^i 
\\[2mm] \nonumber
&& ~~~~~~~~~~~~~~~~
+ \, \omega \sum_\sigma q_\sigma \int \frac{d^3 p}{(2\pi)^3}\,
\frac{v^i} {\omega-\vec v\cdot\vec k}\, 
\bigg(\delta n_{0\sigma}(\vec k,\vec p) 
-i \frac{q_\sigma}{2\omega}\big(\vec v \times \vec B_0(\vec k)\big)\cdot \vec\nabla_p n_\sigma(\vec p)\bigg) \,,
\eea
where 
\bea
\label{full-prop}
\big(\Delta^{-1}(\omega,\vec k)\big)^{ij} = \big(\Delta_{\rm bare}^{-1}(\omega,\vec k)\big)^{ij} - \Pi^{ij}(\omega,\vec k)
\eea 
is the retarded inverse gauge field propagator in the hard loop approximation. If we drop the terms in Eq.~(\ref{elec1}) that depend on the initial conditions we get $(\Delta^{ij}\big(\omega,\vec k)\big)^{-1}\,E^j(\omega,\vec k)=0$ which gives the familiar result that the dispersion equation for the collective modes of the system is obtained from setting the determinant of the inverse propagator to zero.

In an isotropic plasma the function $n_\sigma(\vec p)$ depends on only the magnitude $p\equiv|\vec p|$ which means that $\vec\nabla_p n_\sigma(\vec p) \sim \vec p$, and since $\vec p \parallel \vec v$ the last term in the parentheses on the right side of Eq.~(\ref{elec1}) is zero. In an anisotropic system this term is not identically zero, but it is higher order in the coupling and we neglect it. 
We therefore rewrite Eq.~(\ref{elec1}) as 
\bea
\label{elec3}
E^i(\omega,\vec k) = i \Delta^{ij}(\omega,\vec k) 
\Big[\omega \, E_0^j(\vec k) - (\vec k\times \vec B_0(\vec k))^j  -i  \omega  N_0^j(\vec k;\omega)\Big] ,
\eea
where we have defined 
\bea
\label{N0-defn}
N^j_0(\vec k;\omega) \equiv \sum_\sigma q_\sigma \int \frac{d^3 p}{(2\pi)^3}\,\frac{v^j}{\omega-\vec v\cdot\vec k}\, \,\delta n_{0\sigma}(\vec k,\vec p)  \,.
\eea
Using Faraday's law or the homogeneous Maxwell equation (\ref{max-k-2}) it is straightforward to obtain the corresponding expression for the magnetic field
\bea
\label{magnetic3}
B^i(\omega,\vec k) = \frac{1}{\omega}\epsilon^{ijl}k^j E^l(\omega,\vec k) +\frac{i}{\omega}B_0^i(\vec k) \,.
\eea

The expressions (\ref{elec3}, \ref{magnetic3}) determine the electric and magnetic fields which occur in the plasma as responses to the initial conditions given by $\delta n_{0 \sigma}$, $\vec E_0$ and $\vec B_0$. In the next section we show how to use them in equation (\ref{qhat3}) to obtain the transverse momentum broadening coefficient.

%%%%%%%%%%%%%%%%%%%%%%%%%%%%%%%%%%%%%%%%%%%%%%%%%%%%%%%%
\section{Field correlators}
\label{correlators-section}
%%%%%%%%%%%%%%%%%%%%%%%%%%%%%%%%%%%%%%%%%%%%%%%%%%%%%%%%

Equation (\ref{qhat3}) gives $\hat q$ in terms of the field correlators in coordinate space, which can be written as momentum space correlation functions by Fourier transforming. For example, 
\bea
\label{fftE}
\langle E^i(t_1,\vec r_1)\, E^j(t_2,\vec r_2)\rangle &=& \int^{\infty+i\sigma}_{-\infty+i\sigma} \frac{d\omega_1}{2\pi}\,\int^{\infty+i\sigma}_{-\infty+i\sigma} \frac{d\omega_2}{2\pi} \int \frac{d^3 k_1}{(2\pi)^3}\,\int \frac{d^3 k_2}{(2\pi)^3}  
\\[2mm]
&\times&
e^{-i(\omega_1 t_1 - \vec k_1\cdot \vec r_1)} e^{-i(\omega_2 t_2 - \vec k_2\cdot \vec r_2)} \langle
E^i(\omega_1,\vec k_1)E^j(\omega_2,\vec  k_2)\rangle \,,
\nonumber
\eea
where we have used as before $\vec r_i=\vec r(t_i) = \vec{r}_{i}(0)+ \vec u\, t_i$ with $i =1,\,2$. There is a  similar expression for each of the three other correlators $\langle B^i(t_1,\vec r_1)\, B^j(t_2,\vec r_2)\rangle$, $\langle E^i(t_1,\vec r_1)\, B^j(t_2,\vec r_2)\rangle$ and $\langle B^i(t_1,\vec r_1) \, E^j(t_2,\vec r_2)\rangle$. In this section we derive expressions for the momentum space correlators; $\langle E^i(\omega_1,\vec k_1)E^j(\omega_2,\vec  k_2)\rangle$ and the three other correlators which involve magnetic fields. We follow  the method developed in \cite{Mrowczynski:2008ae}. 

Equations (\ref{elec3}) and (\ref{magnetic3}) can be used to express the momentum space field correlators as sums of terms each of which contains a correlator of two of the initial functions $E_0^j(\vec k)$, $B_0^j(\vec k)$ or $N_0^j(\vec k;\omega)$. There are nine such initial correlators: $\langle E^i_0(\vec k_1) \, E^j_0(\vec k_2)\rangle$,  $\langle E^i_0(\vec k_1) \, B^j_0(\vec k_2)\rangle$, $\langle E^i_0(\vec k_1) \, N^j_0(\vec k_2; \omega_2)\rangle$ {\it etc}.
These initial correlators are calculated at the moment in time ($t=0$) when the parton arrives. We assume that in this initial state the system can be treated as a noninteracting classical plasma which is fundamentally described by the statement that two space time points $(t_1,\vec r_1)$ and $(t_2,\vec r_2)$ are correlated, if there is a particle in the system with velocity that allows it to move between them. Mathematically this means we assume 
\bea
\label{free-tx}
&& \langle \delta n_{\sigma_1}(t_1,\vec r_1,\vec p_1) \, \delta n_{\sigma_2}(t_2,\vec r_2,\vec p_2) \rangle \nonumber\\[4mm]
&& ~~~~~~~~~~ =  \delta_{\sigma_1 \sigma_2} \;(2\pi)^3 \delta^3(\vec p_1-\vec p_2)
\, \delta^3\big((\vec r_1 - \vec v_1 t_1)-(\vec r_2-\vec v_2 t_2)\big) n_{\sigma_1}(\vec p_1)\, .
\eea
We also assume that the plasma particles have no internal degrees of freedom and obey Boltzmann statistics. If the latter assumption is relaxed, the distribution function $n_{\sigma}(\vec p)$ from the r.h.s. of Eq.~(\ref{free-tx}) should be replaced by  $n_{\sigma}(\vec p) \big( 1 \pm n_{\sigma}(\vec p) \big)$ where the upper sign is for bosons and the lower one for fermions.  Since the free system is translationally invariant in time,  we use the usual two sided Fourier transform on the correlation function (\ref{free-tx}) and obtain 
\bea
\label{free-omegak}
&&\langle \delta n_\sigma(\omega_1,\vec k_1,\vec p_1) \, 
\delta n_\sigma(\omega_2,\vec k_2,\vec p_2) \rangle 
\\[2mm] \nn
&&~~ =  \delta_{\sigma_1 \sigma_2} (2\pi)^3\delta^3(\vec p_1-\vec p_2)\;
(2\pi)^3\delta^3(\vec k_1+\vec k_2) \;
2\pi \delta(\omega_1-\vec k_1\cdot\vec v_1)\;
2\pi \delta(\omega_2 + \vec k_2\cdot\vec v_2)  \, n_{\sigma_1}(\vec p_1) \,.
\eea
All other initial state correlation functions will be obtained from  the correlation function (\ref{free-omegak}), as explained below.

Equation (\ref{N0-defn}), which defines $N_0^j(\vec k; \omega)$, can formally be rewritten as
\be
\label{N0}
N^j_0(\vec k;\omega) =  \int \frac{d\omega^\prime}{2\pi} 
\int \frac{d^3p}{(2\pi)^3} \;\frac{v^j}{\omega-\vec v\cdot\vec k}\, 
\sum_\sigma q_\sigma\delta n_\sigma(\omega^\prime,\vec k,\vec p) \,.
\ee
The initial field $\vec E_0(\vec k)$ can be obtained from the two sided Fourier transform of Maxwell's equations and has the form
\bea
\label{E0}
E^i_0(\vec k) &=& \int \frac{d\omega^\prime}{2\pi} E^i(\omega^\prime,\vec k)
= -i \int \frac{d\omega^\prime}{2\pi} \omega^\prime \Delta^{ij}_{\rm bare}(\omega^\prime,\vec k)
j^{j}(\omega^\prime,\vec k) 
\\[2mm] \nn
& =& -i \int \frac{d\omega^\prime}{2\pi} \int\frac{d^3p}{(2\pi)^3} \, \omega^\prime 
\Delta_{\rm bare}^{ij}(\omega^\prime,\vec k) \, v^j \,
\sum_\sigma q_\sigma\delta n_\sigma(\omega^\prime,\vec k,\vec p) \,.
\eea
The corresponding expression for $\vec B_0(\vec k)$ is obtained from Eq. (\ref{E0}) using Faraday's law (\ref{max-k-2}). We obtain
\bea
\label{B0}
B^i_0(\vec k)  &=& \int \frac{d\omega^\prime}{2\pi} B^i(\omega^\prime,\vec k)
= \epsilon^{ijl}k^j  \int \frac{d\omega^\prime}{2\pi} \frac{ E^l(\omega^\prime,\vec k) }{\omega^\prime}
\\[2mm] \nn
& =& -i \,\epsilon^{ijl}k^j   \int \frac{d\omega^\prime}{2\pi} \int\frac{d^3p}{(2\pi)^3} \,
\Delta_{\rm bare}^{lm}(\omega^\prime,\vec k) \, v^m \,
\sum_\sigma q_\sigma\delta n_\sigma(\omega^\prime,\vec k,\vec p) \,.
\eea
We notice that the inverse bare propagator in equations (\ref{E0}, \ref{B0}) was obtained using a two sided Fourier transform, and in Eqs.~(\ref{bare-prop}, \ref{full-prop}) it comes from a one sided transform. However, the retarded propagator in coordinate space vanishes for $t<0$ and therefore its one sided and two sided Fourier transforms are the same. 

The initial state correlators $\langle E^i_0(\vec k_1) \, E^j_0(\vec k_2)\rangle$,  $\langle E^i_0(\vec k_1) \, B^j_0(\vec k_2)\rangle$, $\langle E^i_0(\vec k_1) \, N^j_0(\vec k_2; \omega_2)\rangle$ {\it etc}. can now all be determined from equations (\ref{N0}, \ref{E0}, \ref{B0}) and  the free particle correlation function (\ref{free-omegak}). The momentum space field correlators can then be determined from these nine initial correlators, as described in the first paragraph of this section.

%%%%%%%%%%%%%%%%%%%%%%%%%%%%%%%%%%%%%%%%%%%%%%%%%%%%%%%%
\section{Construction of the integrand}
\label{integrand-section}
%%%%%%%%%%%%%%%%%%%%%%%%%%%%%%%%%%%%%%%%%%%%%%%%%%%%%%%%

Using the field correlators whose derivation is described in the previous section, we can compute the parameter $\hat{q}$ in Eq.~(\ref{qhat3}). The expressions for the correlators of the initial values (\ref{N0}, \ref{E0}, \ref{B0}) involve integrals over $\omega^\prime_1$, $\omega^\prime_2$, $\vec p_1$, and $\vec p_2$. The coordinate-space field correlators are written as integrals over $\omega_1$, $\omega_2$, $\vec k_1$ and $\vec k_2$ of the corresponding momentum space quantities (see Eq. (\ref{fftE})). The delta functions in Eq.~(\ref{free-omegak}) can be used to perform the integrals over $\omega^\prime_1$, $\omega^\prime_2$, $\vec k_2$ and $\vec p_2$. Denoting $\vec k \equiv \vec k_1$ and $\vec p \equiv \vec p_1$, the lengthy result of the entire procedure can be written in the form
\bea
\label{qhat4pre}
&& \hat q = e^2 
\sum_\sigma q_\sigma^2\int^{\infty+i\sigma}_{-\infty+i\sigma} \frac{d\omega_1}{2\pi}
\int^{\infty+i\sigma}_{-\infty+i\sigma} \frac{d\omega_2}{2\pi} 
\int \frac{d^3k}{(2\pi)^3}\int \frac{d^3p}{(2\pi)^3} \, n_\sigma(\vec p) 
\\[3mm]&& ~~~~~~~~~~~~~~
\times \big[ \, {\cal I}_{EE}(t)  \, {\cal C}_{EE} + \, {\cal I}_{EB}(t)  \, {\cal C}_{EB} 
+ \, {\cal I}_{BE}(t)  \, {\cal C}_{BE} + \, {\cal I}_{BB}(t)  \, {\cal C}_{BB}\big] \,.
\nn
\eea
In the square bracket in the last line of Eq.~(\ref{qhat4pre}), we have divided the contributions from the four different correlators. For each correlator, the factor ${\cal I}_{XY}(t)$ (with $\{X,Y\}\in\{E,B\}$) contains all of the time dependence, and in addition depends on $\omega_1$, $\omega_2$, $\vec k$ and $\vec u$. For the electric field correlators, ${\cal I}_{EE}(t)$ is obtained when the field correlators (\ref{fftE}) are  substituted in Eq.~(\ref{qhat3}) and equals
\bea
\label{MBexponentialI}
{\cal I}_{EE}(t) = \frac{d}{dt} \int_0^t dt_1\int_0^t dt_2\;
 e^{-i(\omega_1  -\vec k\cdot\vec u)t_1}  e^{-i(\omega_2  + \vec k\cdot\vec u)t_2} \,.
\eea
The notation ${\cal C}_{EE}$ indicates the contribution from all other factors. 

When the magnetic field enters the correlator, the exponential function $e^{-i\omega_i t_i}$ is replaced by $\big(e^{-i\omega_i t_i}-1\big)$. The factors ${\cal I}_{EB}(t)$, ${\cal I}_{BE}(t)$, and ${\cal I}_{EE}(t)$ are therefore
\bea
\label{EB-exp}
&& {\cal I}_{EB}(t) =\frac{d}{dt} \int_0^t dt_1\int_0^t dt_2
 e^{-i(\omega_1  -\vec k\cdot\vec u)t_1}  \big(e^{-i\omega_2 t_2}-1\big) e^{-i \vec k\cdot\vec u t_2} , 
\\
\label{BE-exp}
 && {\cal I}_{BE}(t) = \frac{d}{dt} \int_0^t dt_1\int_0^t dt_2
 \big(e^{-i\omega_1 t_1}-1\big) e^{ i\vec k\cdot\vec u t_1}  e^{-i(\omega_2  + \vec k\cdot\vec u)t_2} , 
\\
\label{BB-exp}
 && {\cal I}_{BB}(t) =  \frac{d}{dt} \int_0^t dt_1\int_0^t dt_2
 \big(e^{-i\omega_1 t_1}-1\big) e^{ i\vec k\cdot\vec u t_1}  \big(e^{-i\omega_2 t_2}-1\big) e^{-i \vec k\cdot\vec u t_2} .
\eea
Mathematically, the extra $-1$ reduces by one the order of the pole at $\omega =0$, and is necessary to obtain a finite result.
To see why, we note that the correlator involving the magnetic field $B(\omega,\vec k)$ has a pole at $\omega = 0$ which is one order higher than the corresponding correlator with the electric field $E(\omega,\vec k)$. This is evident from the form of Faraday's law (\ref{max-k-2}) which relates the electric and magnetic fields. Physically the introduction of the $-1$ terms in equations (\ref{EB-exp} - \ref{BB-exp}) is necessary to constrain the field solutions to forms which have well defined one sided Fourier transforms. A detailed explanation of these $-1$ terms is given in Appendix \ref{zero-poles-section}. The factors ${\cal C}_{EB}$, ${\cal C}_{BE}$ and ${\cal C}_{BB}$ contain all other contributions from the corresponding correlators.

We note that the result for $\hat q$ in Eq. (\ref{qhat4pre}) is clearly zero in vacuum, since the integrand contains a factor of the distribution function $n_\sigma(\vec p)$. This shows that the interaction of the parton with its own field does not contribute to the momentum broadening coefficient, which is necessary for the consistency of the procedure (see the discussion below equation (\ref{ind-current})).

We can rewrite Eq. (\ref{qhat4pre}) using the same factorization trick as was discussed above equation (\ref{PI3}). In the ultrarelativistic limit we have
\bea
\label{qhat4}
&& \hat q = e^2 d_{\rm scale} 
\int^{\infty+i\sigma}_{-\infty+i\sigma} \frac{d\omega_1}{2\pi}
\int^{\infty+i\sigma}_{-\infty+i\sigma} \frac{d\omega_2}{2\pi} 
\int \frac{d^3k}{(2\pi)^3}\int \frac{d\Omega}{4\pi} 
\\[3mm]&& ~~~~~~~~~~~~~~
\times \big[ \, {\cal I}_{EE}(t)  \, {\cal C}_{EE} + \, {\cal I}_{EB}(t)  \, {\cal C}_{EB} 
+ \, {\cal I}_{BE}(t)  \, {\cal C}_{BE} + \, {\cal I}_{BB}(t)  \, {\cal C}_{BB}\big] \,,
\nn
\eea
where we have defined
\bea
\label{dscale-def}
d_{\rm scale} \equiv \sum_\sigma q_\sigma^2 \int\frac{d^3p}{(2\pi)^3} n_\sigma(\vec p)\,.
\eea
The parameter $d_{\rm scale}$ characterizes the average transverse momenta of the distribution and is discussed in detail in Appendix \ref{mass-def-section}. 

The calculation of the four ${\cal C}$ factors in Eq. (\ref{qhat4}) is straightforward but extremely tedious. We have done it using {\it Mathematica}. The method is described in \cite{Carrington:2013koa} and has been tested in this context by calculating the integrand for the equilibrium plasma (see Appendix \ref{equib-appendix}). As an example, we consider the 8th term in ${\cal C}_{EE}$ which is
\bea
\label{example1}
{\cal C}_{EE}^{(8)} =  \frac{(\hat k\cdot \vec v)^2  (\hat k \cdot \vec u)^2 \,\Delta_A(\omega_1,\vec k)\,\Delta_A(\omega_2,-\vec k)}{(\hat\omega_1 - \vec v\cdot \hat k)(\hat\omega_2+\vec v\cdot \hat k)\big(1-(\vec v\cdot\hat k)^2\big)}\,.
\eea
Substituting the expression (\ref{example1}) into Eq.~(\ref{qhat4}) gives
\bea
\label{example}
&& \hat q^{(8)}_{EE} = e^2 d_{\rm scale}  
\int^{\infty+i\sigma}_{-\infty+i\sigma} \frac{d\omega_1}{2\pi}
\int^{\infty+i\sigma}_{-\infty+i\sigma} \frac{d\omega_2}{2\pi} 
\int \frac{d^3 k}{(2\pi)^3}  \int \frac{d\Omega}{4\pi}  \, {\cal I}_{EE} (t) 
\\[2mm] \nn
&& ~~~~~~~~~~~~~~~~~~~~\times \, 
\Delta_A(\omega_1,\vec k)\Delta_A(\omega_2,-\vec k)\,
\frac{ (\hat k \cdot \vec v)^2 (\hat k \cdot \vec u)^2}
{(\hat\omega_1-\vec v\cdot\hat k)(\hat\omega_2+\vec v\cdot\hat k)\big(1-(\vec v\cdot \hat k)^2\big)} ,
\eea
where $\hat\omega_i \equiv \omega_i/k$ with $k\equiv |\vec k|$ and $\hat k \equiv \vec k/k$. In the rest of this paper, we will explain several aspects of our procedure with reference to this example. 

The integrals over $\omega_1$ and $\omega_2$ will be done by closing the contour in the lower half plane (see equations (\ref{1side-forward}, \ref{1side-inverse}) and the discussion below). In equilibrium plasma (where all collective modes are damped and give contributions exponentially decaying in time) we include only the contributions from the Landau poles obtained from the factors $(\hat\omega_1-\vec v\cdot \hat k)(\hat\omega_2+\vec v\cdot \hat k)$ in the denominator of each term (see Eq. (\ref{example})). These poles give time independent contributions to $\hat q$. The collective excitations of an anisotropic plasma include unstable modes (modes with positive imaginary parts). These unstable modes are crucially important in the calculation of momentum broadening. Due to the factors (\ref{MBexponentialI}, \ref{EB-exp}, \ref{BE-exp}, \ref{BB-exp}), they give contributions to $\hat q$ that grow exponentially in time and overwhelm all other contributions in the long time limit. In the next section we define the anisotropic distribution function that we will use, and describe the dispersion relations it produces.

%%%%%%%%%%%%%%%%%%%%%%%%%%%%%%%%%%%%%%%%%%%%%%%%%%%%%%%%
\section{Extremely oblate plasma}
\label{distro-section}
%%%%%%%%%%%%%%%%%%%%%%%%%%%%%%%%%%%%%%%%%%%%%%%%%%%%%%%%

In this section we introduce the specific anisotropic momentum distribution that we will use in this paper. We define the  momentum distribution and discuss the spectrum of plasmons - collective modes of gauge bosons. We start with a brief discussion of an equilibrium isotropic plasma. 

In equilibrium the distribution of plasma constituents depends only on the magnitude of the momentum $p \equiv |\vec p|$ and can be represented as a sphere in momentum space. The gauge field propagator $\Delta^{ij}(\omega,\vec k)$ can be split into two components commonly denoted $\Delta_T(\omega,\vec k)$ and $\Delta_L(\omega,\vec k)$ using two projection operators which are transverse and longitudinal with respect to the momentum $\vec k$. Solutions of the dispersion equations  $\Delta_T^{-1}(\omega,\vec k) = 0$ and $\Delta^{-1}_L(\omega,\vec k)=0$ give the well known dispersion relations for the transverse and longitudinal modes $\omega_T(k)$ and $\omega_L(k)$, see {\it e.g.} the textbook \cite{lebellac}. 

An anisotropic momentum distribution can be obtained from the isotropic one in a simple way by either squeezing or stretching it in one direction \cite{Romatschke:2003ms}. In studies of heavy ion collisions, one usually takes the direction of deformation to be the beam axis, which we assume to be the $z$-axis. The squeezed and stretched distributions are called, respectively, oblate and prolate. The special cases of extremely oblate and extremely prolate systems are significantly simpler to study mathematically. The distribution functions for these systems have the form
\bea
n_{\rm ex-prolate} (\vec p) &=& \delta(p_T)\, g(p_L) ,
\label{ext-prolate}
\\
n_{\rm ex-oblate}(\vec p) &=& \delta(p_L)\, h(p_T) ,
\label{ext-oblate}
\eea
where we have written $p_L \equiv \vec p\cdot \hat z$ and $p_T \equiv |\vec p-\hat z p_L|$. In the extremely prolate system, the oscillatory behavior of the integrand, which comes from the real modes, is strong enough that the growth produced by the imaginary modes is not clearly seen and the magnitude of $\hat{q}$ is similar to that in equilibrium plasmas. For this reason we consider only the extremely oblate system, which is most relevant to the study of heavy ion collisions where the momentum distribution rapidly becomes oblate due to free streaming \cite{Jas:2007rw}. From now on we refer to the extremely oblate distribution as simply `oblate'. We assume that the distributions of all species of plasma particles are oblate. Our notation for the polarization tensor and gauge boson propagator in an oblate system are given in Appendix \ref{prop-appendix}. Full details are available in our extensive study of collective modes in plasma systems which considers all possible degrees of one dimensional deformation of an isotropic momentum distribution \cite{Carrington:2014bla}. The important points are summarized below.

In an oblate system the propagator is expanded in a four component basis constructed from the momentum vector and the vector which specifies the direction of the deformation. The calculation of $\hat q$ in an oblate system is considerably more difficult than in equilibrium, in part because of the more complex structure of the propagator. There are two components of the propagator, which we call $\Delta_A(\omega,\vec k)$ and $\Delta_G(\omega,\vec k)$, that can be obtained analytically and have a relatively simple structure. However, the dispersion relations are much more complicated and can only be obtained numerically. A crucial difference from the equilibrium system is that the collective excitations of the oblate plasma include unstable modes (modes with positive imaginary parts). The dispersion equation $\Delta^{-1}_A(\omega,\vec k) = 0$ has two real solutions for all values of $\vec k$ which we call $\pm \omega_\alpha$.  For $k<k_A$ there are two imaginary solutions denoted $\pm i\gamma_{\alpha i}$. The threshold wavevector $k_A$ is given in equation (\ref{AG-thres}) below.  The dispersion equation $\Delta^{-1}_G(\omega,\vec k) = 0$ has four real solutions for all values of $\vec k$ which we call $\pm \omega_+$ and $\pm \omega_-$. There are also two imaginary solutions, called $\pm\omega_{-i}$, for $k<k_G$. The threshold values $k_A$ and $k_G$ are
\bea
\label{AG-thres}
&&k_A = \frac{m}{\sqrt{2}}\frac{|x|}{\sqrt{1-x^2}} \,,
\\[2mm]
\label{AG-thres2}
&&k_G = \frac{m}{2} \; {\rm Re} \sqrt{\frac{\sqrt{x^2+4} \left| x\right|
   +x^2-2}{1-x^2}}\,,
\eea
where $x \equiv \cos\theta$ and $\theta$ is the angle between the plasmon's wave vector $\vec k$ and the direction of the anisotropy. The spectrum of plasmons in the oblate system is summarized in  Table~\ref{mode-table}.

\begin{table}[t]
\begin{tabular}{|c | c | l | l|}
\hline 
~~~ dispersion equation ~~~ & ~~~ region~~~  & ~~~~~~ modes  ~~~~~~ \\
\hline
$\Delta_A^{-1}(\omega,\vec k)=0$ & $k>k_A$  & ~~ $\pm \omega_\alpha$   \\
          & $k<k_A$  & ~~ $\pm \omega_\alpha$, ~~ $\pm ~~i\,\gamma_{\alpha i}$   \\[2mm]
$\Delta_G^{-1}(\omega,\vec k)=0$ & $k>k_G$   & ~~ $\pm \omega_+,~~\pm \omega_-$   \\
          & $k<k_G$  & ~~ $\pm \omega_+,~~\pm \omega_-$,~~  $\pm i\,\gamma_{- i}$  ~~ \\ 
\hline
\end{tabular}
\caption{Plasmons in the oblate system. \label{mode-table}}
\end{table}

%%%%%%%%%%%%%%%%%%%%%%%%%%%%%%%%%%%%%%%%%%%%%%%%%%%%%%%%
\section{Calculation of the Momentum Broadening Integral}
\label{integral-section}
%%%%%%%%%%%%%%%%%%%%%%%%%%%%%%%%%%%%%%%%%%%%%%%%%%%%%%%%

As mentioned in Sec.~\ref{distro-section}, the direction of anisotropy is taken to define the $z$-axis of our coordinate system. The momentum  $\vec p$ and velocity  $\vec v \equiv \vec p/p$ of a constituent of extremely oblate plasma lies in the $x$-$y$ plane. The momentum of the collective modes $\vec k$ is chosen to lie without loss of generality in the $x$-$z$ plane. 
We need also to define the vector which gives the velocity of the test parton, which we call $\vec u$. 
The three vectors $\vec k$, $\vec v$ and $\vec u$ are written 
\begin{subequations}
\label{vecSubberx}
\bea
\vec v &=& (\cos\varphi,\sin\varphi,0) \,,
\\
\vec k &=& k\, (\sin\theta,0,\cos\theta) \,,
\\
\vec u &=& (\sin\Theta\cos\phi,\sin\Theta\sin\phi,\cos\Theta)\,.
\eea
\end{subequations}
We use the symbols $x \equiv \cos\theta$, $\hat\omega \equiv\omega/k$ and $\hat k \equiv \vec k/k$ in some equations.

We now discuss how to calculate the integrals in equation (\ref{qhat4}). The first step is to do the $\vec p$ integral, which means calculating the integral over the angle $\varphi$.  Most terms contain factors $(\hat\omega_1 - \vec v\cdot\hat k)$ and $(\hat\omega_2+\vec v\cdot\hat k)$ (see Eq.~(\ref{example})) which have poles that must be handled carefully. Our method is to rearrange the integrand using partial fractioning to remove factors with zeros in the denominators. The result is a large set of terms that do not have poles and can be easily integrated, and three remaining integrals that have relatively simple analytic forms. The method is described in more detail in Appendix \ref{pints-appendix}. The result is that after performing the $\varphi$ integrals, we obtain an expression of the form
\bea
\label{qhat5}
\hat q = e^2 d_{\rm scale} 
\int^{\infty+i\sigma}_{-\infty+i\sigma} \frac{d\omega_1}{2\pi}\,\int^{\infty+i\sigma}_{-\infty+i\sigma} \frac{d\omega_2}{2\pi} \int \frac{d^3k}{(2\pi)^3} \sum_i \bigg[{\cal I}_i(t)\; f_i[\Delta_A,\Delta_G] \; g_i(\omega_1,\omega_2,\vec u,\vec k)\bigg] \,.\nonumber\\
\eea
There are many terms in the sum over the index $i$ in Eq. (\ref{qhat5}). For each term in the sum, ${\cal I}_i (t)$ represents one of the four functions (\ref{MBexponentialI} - \ref{BB-exp}) which carries all of the time dependence and $f_i$ is either: 1, $\Delta_A(\omega_1,\vec k)$, $\Delta_A(\omega_2,-\vec k)$, $\Delta_A(\omega_1,\vec k)\cdot\Delta_A(\omega_2,-\vec k)$, $\Delta_G(\omega_1,\vec k)$, $\Delta_G(\omega_2,-\vec k)$ or $\Delta_G(\omega_1,\vec k)\cdot\Delta_G(\omega_2,-\vec k)$. All other factors are grouped together and denoted $g_i$. 

The expression represented in Eq. (\ref{qhat5}) is extremely lengthy. There are contributions from both $A$ and $G$ modes. 
Since the unstable $A$-modes are stronger than the unstable $G$-modes \cite{Carrington:2014bla}, we expect they will give the dominant contribution to $\hat q$. In our calculation of collisional energy loss \cite{Carrington:2015xca}, which is similar in structure to the $\hat q$ calculation, we found that $A$-modes dominate over $G$-modes. 
In addition, the $A$-mode terms are much easier to calculate because the corresponding part of the propagator has a simpler tensor structure (see Appendix~\ref{prop-appendix}).
We have therefore done most of our calculations including only $A$-modes. We have calculated the contribution from the $G$-modes for one choice of the external parameter $\Theta$ and verfied that their contribution is much smaller than the $A$-mode piece. 

There is one tricky point that arises when the integral over azimuthal angle $\varphi$ is done which we discuss below. 
Using our coordinate system (\ref{vecSubberx}), the factor $1-(\vec v\cdot \hat k)^2$, which appears in the denominator of many terms (see Eq.~(\ref{example})), has the form
\bea
\label{C-def}
C(\varphi) = \frac{1}{1-(1-x^2)\cos^2\varphi}\,.
\eea
Many of the integrals we need contain this factor and diverge when $x=0$. It is easy to see that the factor $C(\varphi)$ is produced when the free correlation functions (\ref{free-omegak}) are used in the bare propagators (\ref{bare-prop}), which are part of the initial fields given in Eqs.~(\ref{E0}, \ref{B0}). The divergence is caused by the approximation that the plasma particles are massless, or equivalently the approximation $|\vec v|=1$. Physically it is regulated by the small but non-zero mass of the plasma particles, which do not have the dispersion relations of massless particles in a physical plasma (for example, in equilibrium their masses can be identified with their energies at zero momentum and are of order $gT$, see {\it e.g.} \cite{lebellac}). We introduce a parameter $m_{\rm min}$ in the denominator of Eq.~(\ref{C-def}) (see equation (\ref{generic-phi1-b})), and in Sec.~\ref{results-section} we show that the dependence of the momentum broadening coefficient on this parameter is logarithmic. The parameter $m_{\rm min}$ is not determined by our formalism and must be introduced by hand, which is clearly a weakness of our approach. 
However, it is also clear that this divergence is not directly related to the effect we are looking for. Firstly, it enters with the initial conditions and does not depend on the properties of the distribution function. Secondly, the region of the momentum space which is affected is $x\to 0$, and from the formulae (\ref{AG-thres}, \ref{AG-thres2}) and the discussion in Sec.~\ref{distro-section} we see that the unstable modes disappear when $x=0$. We conclude therefore that although the result for $\hat q$ does depend weakly on the value of $m_{\rm min}$, the exponential increase in $\hat q$ as a function of time is a physical effect that is not related to the introduction of the $m_{\rm min}$ regulator. 

The last part of the calculation that can be done analytically is the frequency integrals. We include only the contributions from the poles of the propagators. The pole structure of the oblate system is briefly discussed in Sec.~\ref{distro-section}. The functions $\Delta_A$ and $\Delta_G$ are known analytically, but we have only numerical expressions for the dispersion relations. 
The functions $\Delta_A$ and $\Delta_G$ have simple poles at the solutions listed in Table \ref{mode-table}, and therefore we calculate the frequency integrals by writing 
\bea
\Delta(\omega,\vec k) = \sum_i \frac{Z(\omega_i,\vec k)}{\omega-\omega_i} \,,
\eea
where
\bea
Z^{-1}(\omega_i,\vec k) = \frac{d}{d\omega}\Delta^{-1}(\omega,\vec k)\bigg|_{\omega=\omega_i} \;.
\eea

After performing the frequency integrals we obtain an expression which depends on the external variables $t$, $\Theta$ and the integration variables $k$, $x$ and $\varphi$. The integrals over $k$, $x$ and $\varphi$ are done numerically. The integrand is even in $x$ and therefore we only need to calculate the $x$ integral from 0 to 1. It grows rapidly when $x\to 1$ because of the influence of the unstable mode which dominates at $x$ close to unity (see Eqs.~(\ref{AG-thres}, \ref{AG-thres2})). The calculation is therefore done most efficiently using a logarithmic variable $x_L$ which is defined as $x_L \equiv \ln (1-x)$. 

The momentum broadening coefficient increases logarithmically with the upper limit of the $k$ integral, which we call $k_{\rm max}$.  The parameter $k_{\rm max}$ is a separation scale which divides the momentum range into two pieces which are relevant to the soft and hard contributions, and should cancel when they are combined. In case of the equilibrium plasma the problem was studied in detail in \cite{Arnold:2008vd}. An investigation of this cancellation in unstable plasma is beyond the scope of this work.

%%%%%%%%%%%%%%%%%%%%%%%%%%%%%%%%%%%%%%%%%%%%%%%%%%%
\section{Quark-gluon plasma}
\label{QGP-section}
%%%%%%%%%%%%%%%%%%%%%%%%%%%%%%%%%%%%%%%%%%%%%%%%%%%

Our formalism applies to ultrarelativistic QED and QCD plasmas, which is a strength of the method that should be emphasized. In the body of this paper we have used language that is applicable to an electromagnetic plasma. In this section we discuss how to modify our expressions so that they apply to QCD plasma. The basic idea is that one must specify which quantities carry color indices, and how to calculate the corresponding group factors. 

The starting point of our approach applied to QGP are the Wong equations \cite{Wong:1970fu} which describe a classical test parton interacting with the chromodynamic field present in the plasma. The Wong equations are usually written in the Lorentz covariant form 
\bea
\label{EOM-1b}
\frac{d p^\mu(\tau)}{d \tau} &=& g Q^a(\tau ) \, F_a^{\mu \nu}\big(x(\tau )\big) \, u_\nu(\tau ) \,,
\\
\label{EOM-1c}
\frac{d Q_a(\tau)}{d \tau} &=& - g f^{abc} u_\mu (\tau ) \, A^\mu _b \big(x(\tau )\big) \, Q_c(\tau) \,,
\eea
where $\tau$, $x^\mu(\tau )$, $u^\mu(\tau)$ and  $p^\mu(\tau)$ are, respectively, the parton's  proper time, trajectory, four-velocity and  four-momentum; $F_a^{\mu \nu}$ and $A_a^\mu$ denote, respectively, the chromodynamic field strength tensor and four-potential in the adjoint representation of the ${\rm SU}(N_c)$ gauge group with the color index $a = 1, \; 2, \dots N_c^2 -1$; $g$ is the coupling constant, which is assumed to be small, $f^{abc}$ is the structure constant of  the ${\rm SU}(N_c)$ group, and finally $gQ^a$ is the classical color charge of the parton.

In order to solve the Wong equations, we adopt two simplifying assumptions. In Sec.~\ref{problem-section} we have already discussed the requirement that the parton's velocity is a unit constant vector.  Now we choose in addition the gauge condition
\bea
\label{gaugeCondition}
u_\mu (\tau ) \, A^\mu _a \big(x(\tau )\big) = 0 \,,
\eea
which makes the potential vanish along the parton's trajectory. Applying the condition (\ref{gaugeCondition}), the second Wong equation (\ref{EOM-1c}) simply states that the parton's charge is a constant of motion, or that $Q_a$ is independent of $\tau$. 

The first Wong equation (\ref{EOM-1b}) can be solved to obtain an expression of the form (\ref{p-soln}) and, repeating the rest of the steps described in  Sec.~\ref{problem-section}, we obtain a formula for the momentum broadening parameter of the form
\bea
\label{qhat-QGP-1}
\hat q_{\rm color} =  Q_a Q_b \,\frac{d}{dt} \int_0^t dt_1 \int_0^t dt_2 
\langle \vec F_a (t_1,\vec r_1 ) \cdot \vec F_b (t_2,\vec r_2) 
- \vec u \cdot \vec F_a(t_1,\vec r_1) \,\vec u \cdot \vec F_b(t_2,\vec r_2) \rangle  \,,
\eea
where $\vec F_a (t,\vec{r}) \equiv g\big( \vec{E}_a (t,\vec{r}) + \vec{u} \times \vec{B}_a(t,\vec{r}) \big)$ and, as previously,  $\vec r_i \equiv \vec r(t_i) = \vec{r}_i(0) + \vec u t_i$ with $i = 1,\,2$. The subscript `color' indicates that the parameter  depends on the color charge of the test parton. Color is not an observable quantity however, and $\hat q_{\rm color}$ is not gauge invariant. In order to obtain a gauge invariant observable, the quantity $\hat q_{\rm color}$ is averaged over the parton's colors using the relation \cite{Litim:2001db}
\be
\label{ave-color-charge-2}
\int dQ \,Q_a Q_b  = C_2  \delta^{ab},
\ee
where $C_2=1/2$ for a quark in the fundamental representation of the ${\rm SU}(N_c)$ gauge group and $C_2=N_c$ for a gluon in the adjoint representation. The momentum broadening parameter averaged over colors is 
\be
\label{qhat-color-ave}
\hat q \equiv \left\{
\begin{array}{ccl}
\frac{1}{N_c}\int dQ \;\hat q_{\rm color}
&{\rm for} & {\rm a~quark} 
\\[3mm]
\frac{1}{N_c^2-1}\int dQ \;\hat q_{\rm color}  &{\rm for}  & {\rm a~gluon}
\end{array}
\right.
\ee
and using Eq. (\ref{ave-color-charge-2}) we obtain 
\bea
\label{qhat-QGP-2}
\hat q =  \frac{C_R}{N_c^2-1} \, \frac{d}{dt} \int_0^t dt_1 \int_0^t dt_2 
\langle \vec F_a (t_1,\vec r_1 ) \cdot \vec F_a (t_2,\vec r_2) 
- \vec u \cdot \vec F_a(t_1,\vec r_1) \,\vec u \cdot \vec F_a(t_2,\vec r_2) \rangle  ,
\eea
with the color factor $C_R$ given by
\begin{displaymath}
C_R \equiv \left\{
\begin{array}{ccl}
\frac{N_c^2-1}{2N_c}
&{\rm for} & {\rm a~quark} 
\\[3mm]
N_c &{\rm for}  & {\rm a~gluon \,.}
\end{array}
\right.  
\end{displaymath}

The field correlators  $\langle E^i_a(t_1,\vec r_1)\, E^j_b(t_2,\vec r_2)\rangle$, $\langle E^i_a(t_1,\vec r_1)\, B^j_b(t_2,\vec r_2)\rangle$ {\it etc.} are found from the linearized Yang-Mills equations which, in a non-covariant three-vector notation, have the familiar form of Maxwell equations (\ref{max-coord-homo}, \ref{max-coord-inhomo}). However, in the QCD calculation the fields $\vec E_a$ and $\vec B_a$ and the sources $\rho_a$ and $\vec j_a$ carry color indices and are chosen to belong to the adjoint representation of the ${\rm SU}(N_c)$ gauge group.
If one considers a plasma which on average is locally color neutral, the correlators are of the form
\be
\label{EE-form-QCD}
\langle E^i_a(t_1,\vec r_1)\, E^j_b(t_2,\vec r_2)\rangle = \delta_{ab} \langle E^i (t_1,\vec r_1)\, E^j(t_2,\vec r_2)\rangle_{\rm EM}
\ee 
where $\langle E^i (t_1,\vec r_1)\, E^j(t_2,\vec r_2)\rangle_{\rm EM}$ denotes a correlator of two electric fields which has the same form as the QED correlator, but with the distribution function defined differently (this point is explained in Appendix \ref{mass-def-section}). Substituting the field correlators of the form (\ref{EE-form-QCD}) into the formula (\ref{qhat-QGP-2}), the momentum broadening parameter is rewritten as
\bea
\label{qhat-QGP-3}
\hat q =  C_R \, \frac{d}{dt} \int_0^t dt_1 \int_0^t dt_2 
\langle \vec F(t_1,\vec r_1 ) \cdot \vec F (t_2,\vec r_2) 
- \vec u \cdot \vec F(t_1,\vec r_1) \,\vec u \cdot \vec F(t_2,\vec r_2) \rangle_{\rm EM}  \,,
\eea
where the trace $\delta_{aa} = N_c^2-1$ is taken into account.

It is not obvious that the QCD momentum broadening parameter (\ref{qhat-QGP-3}) is gauge invariant. In electromagnetism the fields themselves are gauge invariant and therefore so are their correlation functions. The entire calculation of the momentum broadening parameter is therefore manifestly gauge invariant. In QCD however, the fields are gauge dependent and in general their correlators, which are non-local in space-time, change in a complicated way under gauge transformations. The classical color charges also vary under gauge transformations. However, our result for the momentum broadening parameter is in fact gauge invariant. This is explained below.

To prove that the formula for $\hat{q}$ given by Eq.~(\ref{qhat-QGP-3}) is gauge invariant, it is sufficient to prove the invariance of the expression 
\be
W \equiv \int dQ \, Q_a(x) \, Q_b(x') \langle H_a(x) \, H_b (x') \rangle \, ,
\ee
where $Q_a(x)$ is a classical color charge and $H_a(x)$ is a component of chromoelectric or chromomagnetic field. 
The dependence on $x$ is assigned not only to the fields but to the color charge as well, because  $Q_a(x)$ and $H_a(x)$ are subjects of gauge transformations, which are local in space-time. We consider what happens to the quantity $W$ when $Q_a(x)$ and $H_a(x)$ are transformed as
\be
Q_a(x) \to Q_a(x) + f^{abc} Q_b(x) \, \lambda_c(x) \,, 
~~~~~
H_a(x) \to H_a(x) + f^{abc} H_b(x) \, \lambda_c(x) \, ,
\ee
where $\lambda_c(x)$ is an infinitesimal transformation parameter. As the integration measure $dQ$ is gauge invariant  \cite{Litim:2001db}, the quantity $W$ is changed by 
\bea
\nn
\delta W &&=  \int dQ \Big[ f^{acd} Q_d(x) \, \lambda_d(x) \, Q_b(x') \langle H_a(x) \, H_b (x') \rangle
+ Q_a(x)  f^{bcd} Q_c(x')  \, \lambda_d(x') \langle H_a(x) \, H_b (x')  \rangle
\\[2mm] \nn
&& 
%~~~~~~~~~~~~~~ 
+ Q_a(x) \, Q_b(x')   f^{acd} \lambda_d(x) \langle H_c(x) \, H_b (x') \rangle
+  Q_a(x) \, Q_b(x')   f^{acd} \lambda_d(x') \langle H_a(x) \, H_d (x') \rangle \Big] ,\nn\\
\label{del-W}
\eea
where only terms linear in $\lambda$ are kept. All field correlators are unit matrices in color space (see Eq.~(\ref{EE-form-QCD})). This result is basically a consequence of the linear response approximation in which the fields are assumed to be small fluctuations around a color neutral state. Performing the integration over color charge according to Eq.~(\ref{ave-color-charge-2}) and using $\langle H_a(x) \, H_b (x') \rangle\sim\delta_{ab}$, one finds that in every term in Eq.~(\ref{del-W}) the structure constant appears in the form $f^{aab}$ and therefore gives zero because of the anti-symmetric character of these constants. The result is that $\delta W$ vanishes identically, and therefore the momentum broadening parameter (\ref{qhat-QGP-3}) is gauge independent. 

The conclusion is that the integrand that gives $\hat q$ in a QCD plasma has exactly the same form as our result for a QED plasma in Eq. (\ref{qhat5}), with the factor $e^2$ replaced by $g^2 C_R$, and a different definition of the dimensional parameter $d_{\rm scale}$ (see Appendix \ref{mass-def-section}).

%%%%%%%%%%%%%%%%%%%%%%%%%%%%%%%%%%%%%%%%%%%%%%%%%%%
\section{Results}
\label{results-section}
%%%%%%%%%%%%%%%%%%%%%%%%%%%%%%%%%%%%%%%%%%%%%%%%%%%

Hard jets are produced in relativistic heavy-ion collisions at the instant of collision, together with numerous softer partons which constitute a plasma medium that the jets travel through. The momentum distribution of the plasma constituents is initially prolate -- elongated along the beam -- but due to  longitudinal expansion \cite{Jas:2007rw} it becomes oblate after a short period of time, and further evolves towards isotropy. The plasma is unstable both in the prolate and oblate phase. However, we have found that the effect of the imaginary modes on $\hat{q}$ is not clearly seen in the extremely prolate plasma. This happens because of the oscillatory behavior of the integrand, which comes from the contributions from the real modes. Our calculations show that the magnitude of $\hat{q}$ is similar to the equilibrium value, and therefore the short prolate phase cannot much influence the total momentum broadening of the test parton. For this reason, we have focused on the extremely oblate plasma. In this section we present our numerical results.

\begin{figure}[t]
\center
\includegraphics[width=0.9\textwidth]{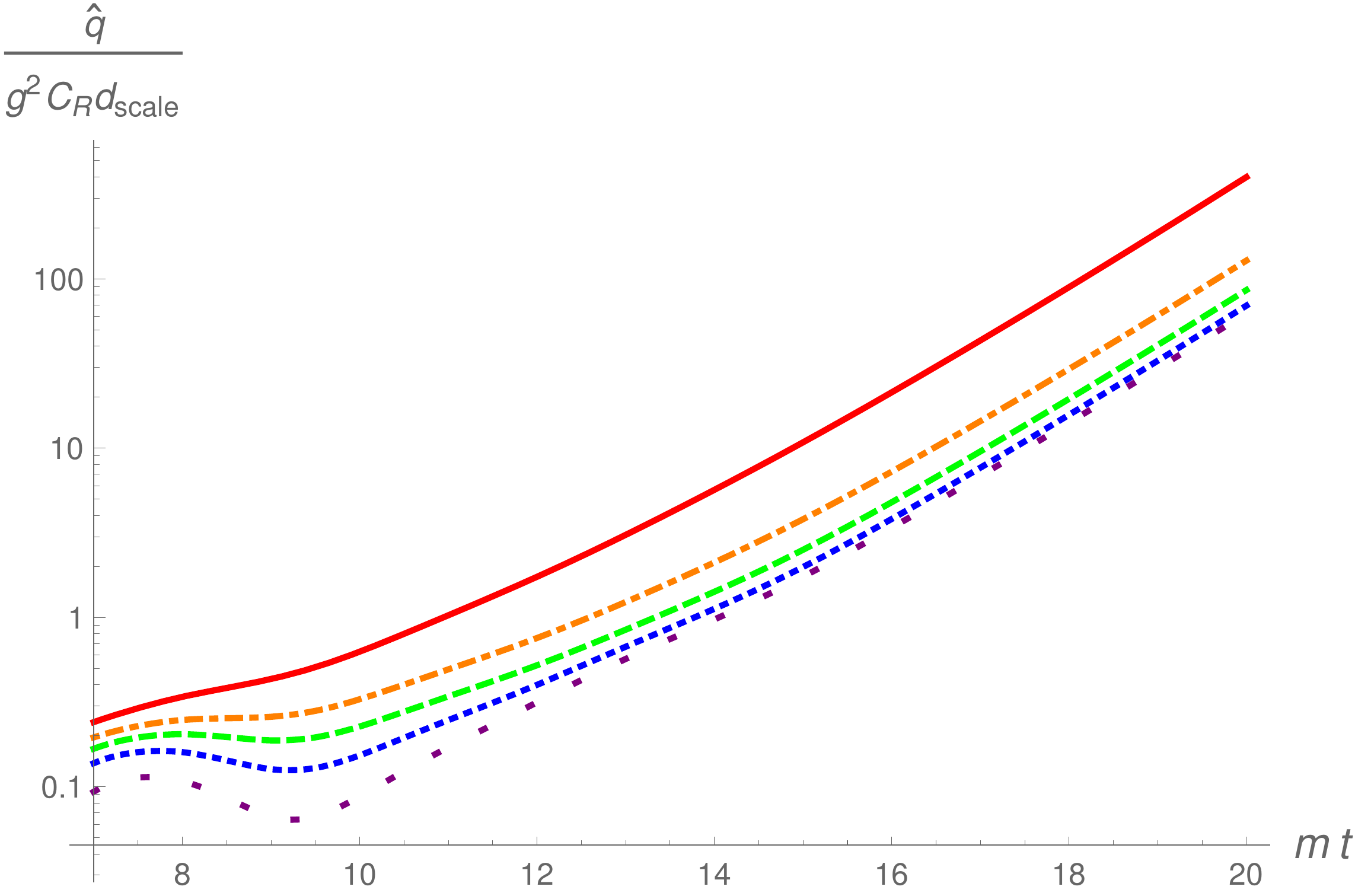}
\caption{(color online) Momentum broadening $\hat q$ as a function of time for five angles $\Theta$ between the test parton velocity and the direction of the anisotropy. The values of $\Theta$ for each line are $\pi/2$ (red, solid); $\pi/3$  (orange, dot-dashed); $\pi/4$ (green, dashed); $\pi/6$ (blue, dotted); $\pi/12$  (purple, spaced dots). For comparison, the equilibrium result is $\hat q = 0.11 \,g^2 C_R d_{\rm scale}$.}
\label{qhatversustime}
\end{figure}

Figure~\ref{qhatversustime}  shows the momentum broadening parameter $\hat q$ as a function of time for five different angles $\Theta$ between the test parton velocity and the $z$-axis, along which the plasma momentum distribution is infinitely squeezed. We use $m_{\rm min}=5\times 10^{-4}\,m$ and $k_{\rm max}=2\,m$, and the value of $\hat q$ is scaled by $g^2 C_R d_{\rm scale}$. 

\begin{figure}[b]
\center
\includegraphics[width=0.5\textwidth]{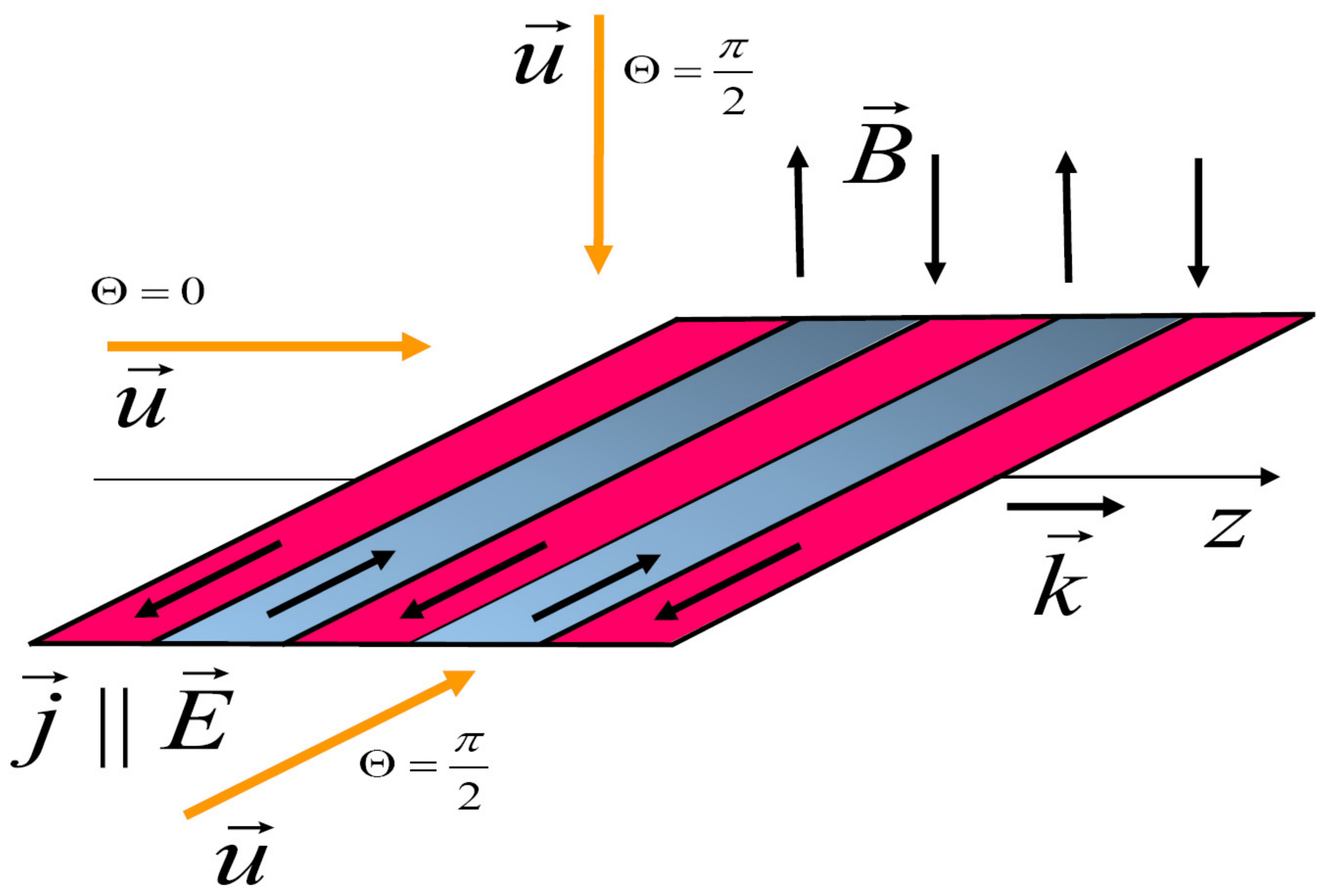}
\caption{(color online) The orientation of the wave vector $\vec k$,  current $\vec j$, electric $\vec E$ and magnetic $\vec B$ fields of the fastest unstable filamentation mode in the oblate plasma. The beam axis $z$ and three orientations of the test parton velocity $\vec u$ are also shown. Momentum broadening is smallest when the parton moves along the $z$-axis ($\Theta =0$), and greatest when it moves perpendicular to $\hat z$ ($\Theta =\pi/2$).}
\label{fig-config}
\end{figure}

The parameter $\hat q$ in oblate plasma should be compared to the momentum broadening $\hat q_{\rm eq}$ in the equilibrium isotropic system of the same mass parameter $m$. For $k_{\rm max}=2m$, we have  $\hat q_{\rm eq}= 0.11 \, g^2 C_R d_{\rm scale}$ (see Appendix \ref{equib-appendix}). The momentum broadening is close to its equilibrium value for short times $t \lesssim 10/m$. At later times we observe the exponential growth of $\hat q$ to values that much exceed $\hat q_{\rm eq}$. The delay in the onset of exponential growth is caused by the fact that in our approach the test parton enters the plasma at the moment when the unstable modes are initiated. The effect of the exponentially growing gauge fields becomes important when their amplitudes start to exceed their typical equilibrium value ($\sim gT$). At this point, the broadening of the momentum of the test parton becomes dominated by its interaction with the unstable modes. 

The curves in Fig.~\ref{qhatversustime} are extended to long times, and thus to rather unrealistically large values of $\hat{q}$. One should remember that our approach is based on the hard loop approximation which requires a separation of soft and hard scales, and therefore breaks down when the amplitude of the exponentially growing mode becomes comparable to that of the hard modes. Consequently, our results are reliable for times which are not too long. This point is addressed in more detail in the concluding section~\ref{sec-conclusions}.

We also observe in Fig.~\ref{qhatversustime} that for all times $\hat q$ is maximal for $\Theta = \pi/2$ and decreases when the angle $\Theta$ tends to zero. This behavior can be understood from the fact that the wave vector $\vec k$ of the fastest mode of the filamentation instability is along the $z$-axis. To see this we note that using our coordinate system (\ref{vecSubberx}) the plasmon wave vector is aligned with the direction of the anisotropy when $\theta=0$ or $x \equiv \cos\theta=1$, and Eqs.~(\ref{AG-thres}, \ref{AG-thres2}) show that the thresholds for unstable modes diverge as $x\to 1$. The strongest $\vec E$ and $\vec B$ fields therefore lie in the $x$-$y$ plane. In the extremely oblate system, when the unstable mode develops, the currents form filaments of charge moving in opposite directions. This is indicated by the alternating strips of pink and blue in Fig. \ref{fig-config}.  (An elementary physics explanation of the mechanism of the instability can be found in \cite{Mrowczynski:1996vh}.) If the test parton enters the plasma along the $z$-axis, it moves through oscillating fields and the overall effect of its interactions with these fields is somewhat weakened. In contrast, if the parton enters perpendicular to $\hat z$, or completely within the $x$-$y$ plane, it interacts with $\vec E$ and $\vec B$ fields of fixed orientation and the change in its momentum is maximal. We also note that the results presented in Fig.~\ref{qhatversustime} assume that the plasma system under consideration is infinite and that the current filaments extend to infinity. In reality one expects that the unstable system is split into domains of a finite size where the filaments are oriented somewhat differently.  

It is interesting to try to quantify the growth of the parameter $\hat q$. The behavior of an unstable system is usually driven by the fastest mode, which in case of an extremely oblate plasma is the pure imaginary mode $\gamma_{\alpha i}$. For $k_{\rm max}=2 m$ its maximal value is found numerically to be $\gamma_{\rm max} = 0.47\, m$. The naive expectation is that $\hat q$ should grow like $e^{2 \gamma_{\rm max}t}$, since this is the growth rate given by the factor ${\cal I}_{EE}(t)$ in Eq.~(\ref{MBexponentialI}). However, the fit of $\hat q$ as a function of time  for $\Theta=\pi/4$ gives the exponent $0.44 \, m$, or approximately the value of $\gamma_{\rm max}$ and not $2\gamma_{\rm max}$.

\begin{figure}[t]
\center
\includegraphics[width=0.7\textwidth]{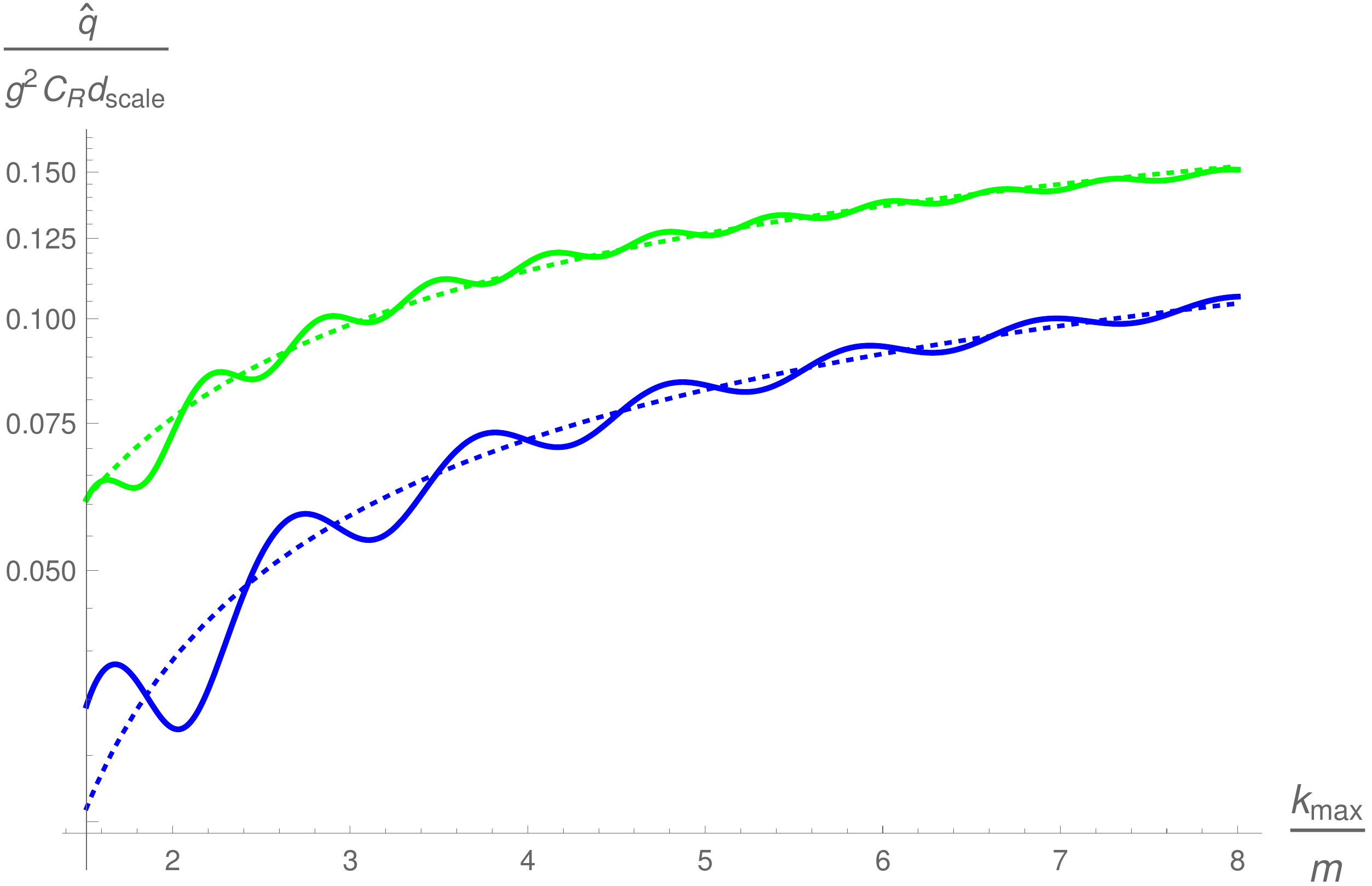}
\caption{(color online) The dependence of $\hat q$ on the scale parameter $k_{\rm max}$. The upper (green) lines are for $t=10/m$ and $\Theta=\pi/6$, and the lower (blue) lines are for $t=6/m $ and $\Theta=\pi/6$. For both times, the dotted line represents a fit of $\hat q$ as a function of $k_{\rm max}$ with $\ln (k_{\rm max}/m)$.}
\label{fig-kmax-dependence}
\end{figure}

In order to understand this behavior better, we have divided the integrand in Eq.~(\ref{qhat4}) into three pieces which correspond to contributions with two, one and zero powers of $N_0$ (which is related to the fluctuation of the distribution function for plasma particles -  see Eq.~(\ref{N0-defn})). We find that terms with two powers of $N_0$ give a positive contribution growing as $e^{0.79 \, m\, t}$, but terms with one power provide a negative contribution with approximately the same exponent. There are therefore large cancellations between these two contributions. The terms with no factors of $N_0$, which depend on the squares of the initial fields, do not grow as rapidly, but they play a role that is larger than expected because of the large cancellations between the terms that depend on $N_0$. The conclusion is that there are pieces of the integrand that grow at double the growth rate of the fastest unstable mode, as expected, but because of large cancellations the net growth is much smaller. 

Our results depend on two regulators $k_{\rm max}$ and $m_{\rm min}$ which signal the incompleteness of our approach. 
We start with a discussion of $k_{\rm max}$. This scale divides the range of momenta transfered to the test parton into two pieces which correspond to the soft and hard contributions. The parameter $k_{\rm max}$ survives in our final results because we have taken into account only the soft piece. In the equilibrium computation the parameter $k_{\rm max}$ disappears when the soft contribution to  $\hat q_{\rm eq}$ is combined with the hard one, which describes elastic collisions between the test parton and plasma constituents with momentum transfer much exceeding the Debye mass. Some details of this cancellation are discussed in \cite{Arnold:2008vd}. It is beyond the scope of the present study to compute the contribution to $\hat q$ from hard scattering in unstable plasma, using a formalism which correctly treats the evolution of the system from its initial conditions. 
We have checked that  $\hat q$ depends on $k_{\rm max}$ logarithmically. In Fig.~\ref{fig-kmax-dependence} we show $\hat q$ as a function of $k_{\rm max}$ for two different times. One observes a mild oscillatory behavior (oscillations are stronger at earlier times), but the general trend is logarithmic. 

\begin{figure}[t]
\center
\includegraphics[width=0.7\textwidth]{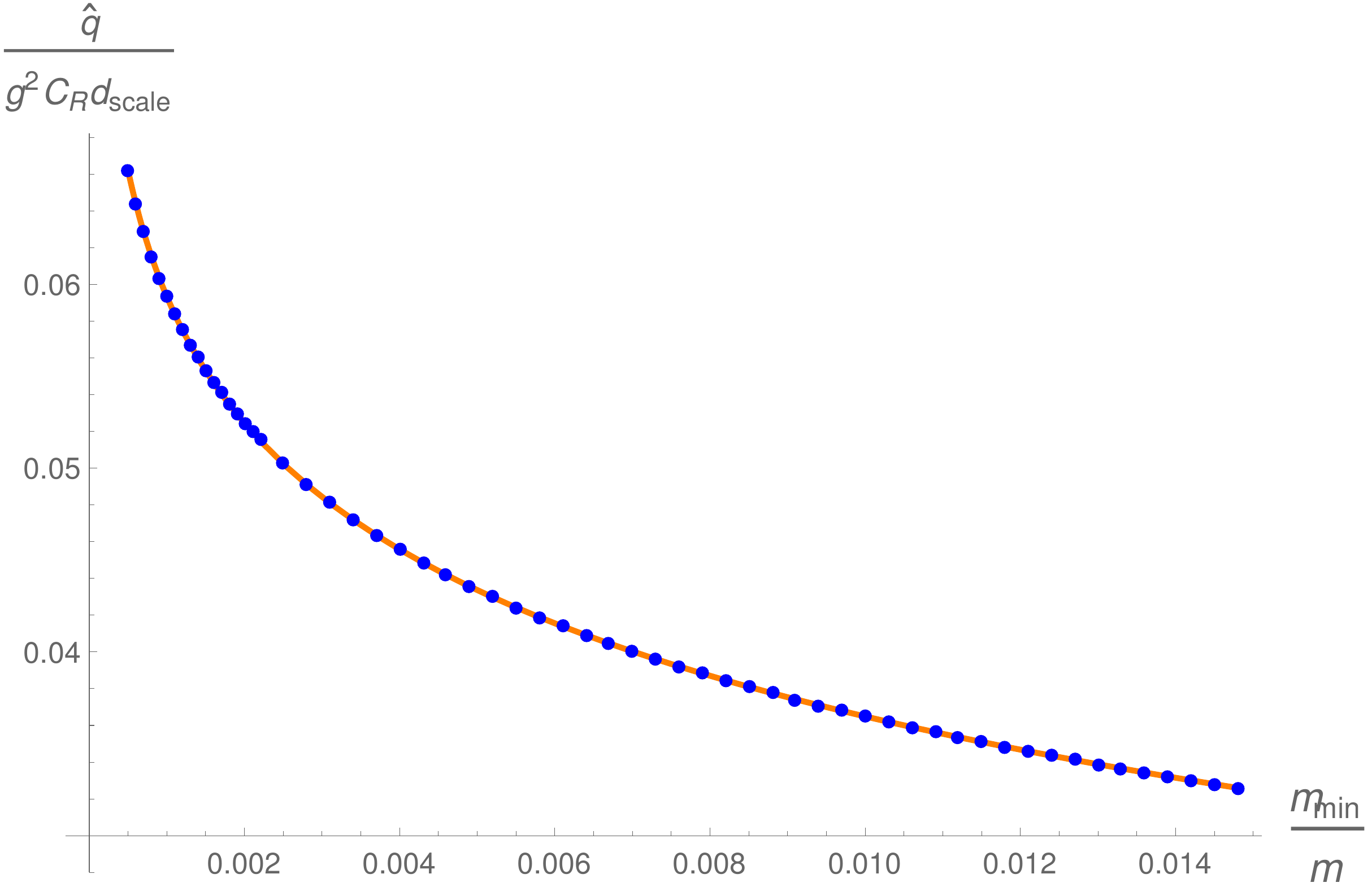}
\caption{(color online) The dependence of $\hat q$ on the scale $m_{\rm min}$, with $t=6/m$ and $\Theta=\pi/6$. The solid (orange) line is a fit of $\hat q$ as a function of $m_{\rm min}$ with $\ln(m/m_{\rm min})$.}
\label{fig-mmin-dependence}
\end{figure}

As discussed in Sec.~\ref{integral-section}, the parameter $m_{\rm min}$ is introduced as a regulator and corresponds physically to the small but non-zero mass of the plasma particles, which was neglected in our formalism. In Fig. \ref{fig-mmin-dependence} we show the dependence of the momentum broadening parameter on the scale $m_{\rm min}$ and a fit of $\hat q$ as a function of $m_{\rm min}$ with ln$(m/m_{\rm min})$. One sees from the graph that the dependence is logarithmic.

\begin{figure}[t]
\center
\includegraphics[width=0.6\textwidth]{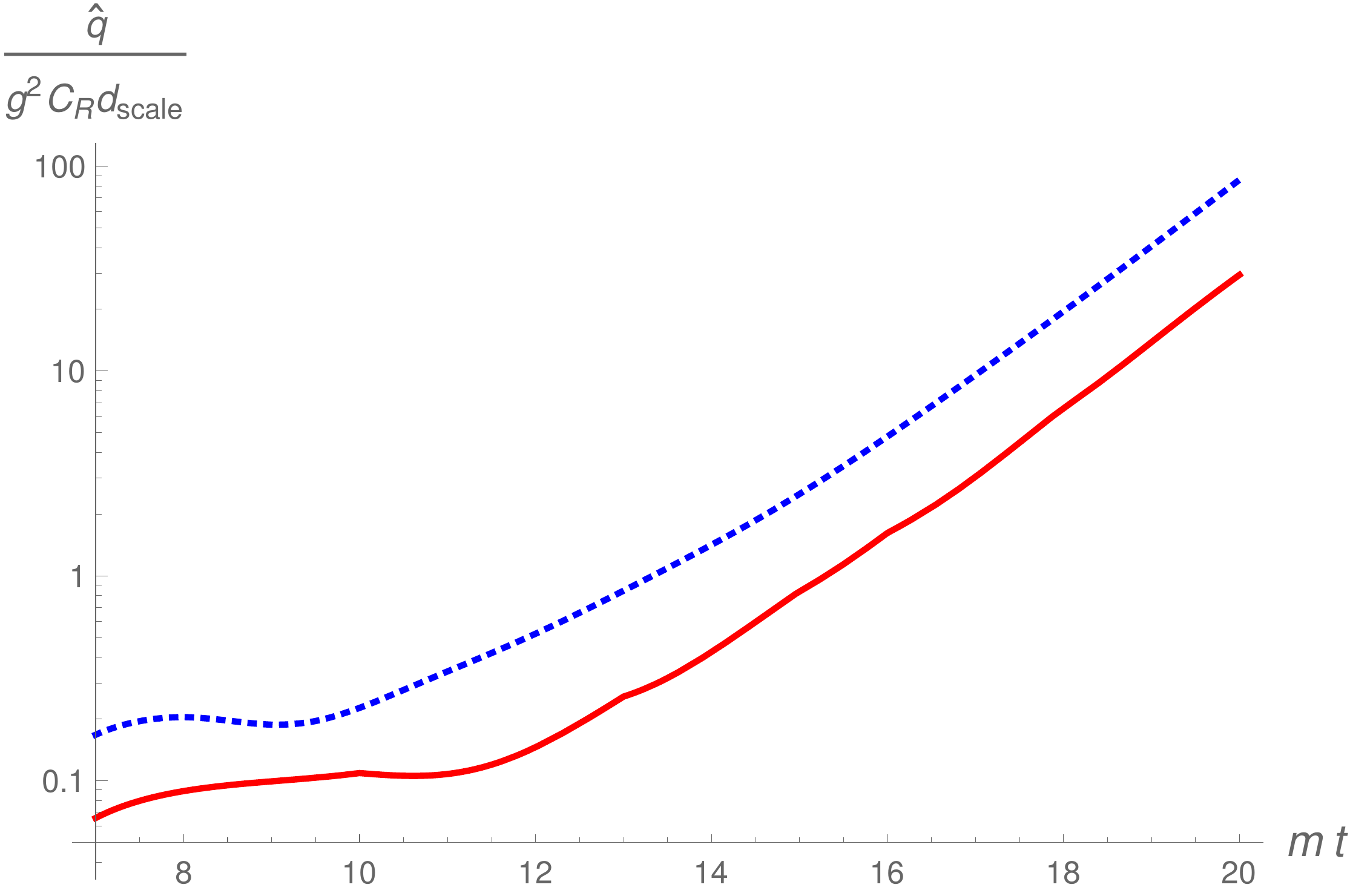}
\caption{(color online) The relative contribution of the $A$-modes (blue, dotted line) and $G$-modes (red, solid line)  for $\Theta=\pi/4$.}
\label{fig-Gmodes}
\end{figure}

As argued at the end of Sec.~\ref{integral-section}, the dominant contribution to $\hat q$ comes from the $A$-modes and for this reason only $A$-modes have been taken into account to obtain the results presented in Fig.~\ref{qhatversustime}-\ref{fig-mmin-dependence}. In Fig. \ref{fig-Gmodes} we show the contributions to $\hat q$ from $A$-modes and $G$-modes with $\Theta=\pi/4$. The figure indeed shows the dominance of the $A$-modes.

%%%%%%%%%%%%%%%%%%%%%%%%%%%%%%%%%%%%%%%%%%%%%%%%%%%%%%%%
\section{Discussion and Conclusions}
\label{sec-conclusions}
%%%%%%%%%%%%%%%%%%%%%%%%%%%%%%%%%%%%%%%%%%%%%%%%%%%%%%%

We start with a discussion of the validity of our results, which are obtained under rather restrictive conditions. We use the hard loop approach which assumes a clear separation of the hard and soft scales. Plasma constituents carry hard momenta and fields are soft which justifies a classical treatment of the fields. The scale separation requires the smallness of the coupling constant which is a crucial limitation of our analysis. 

The equilibrium QGP becomes weakly coupled if its temperature $T$ is much bigger than the QCD scale parameter $\Lambda_{\rm QCD} \sim 200 \; {\rm MeV}$, but the temperature of QGP from relativistic heavy-ion collisions is comparable to $\Lambda_{\rm QCD}$ even at the LHC. However, we deal with the earliest non-equilibrium phase of matter produced in relativistic heavy-ion collisions. In this phase the energy density $\varepsilon$ is much larger than in equilibrium, and the weak coupling condition $\varepsilon^{1/4} \gg \Lambda_{\rm QCD}$ can be satisfied. In central collisions of nuclei of mass number $A$ at the center of mass energy $\sqrt{s}$ per nucleon-nucleon pair, we estimate the initial energy density in the center of mass frame as 
\be
\label{ini-en-den}
\varepsilon = \frac{c_{\rm inel} A \sqrt{s}}{\pi R^2 l}\,,
\ee
where $c_{\rm inel}$ is the inelasticity coefficient -- the fraction of initial energy which goes to particle production, $R$ is the radius of colliding nuclei and $l$ is the length of the cylinder where the energy is released. Assuming that  $c_{\rm inel}=0.5$  \cite{Navarra:2003am} and taking $A=200$, $R = 7$ fm and $l = 1$ fm, one obtains $\varepsilon \approx 3.25 \; {\rm TeV/fm^3}$ for $\sqrt{s}$ = 5 TeV. This corresponds to $\varepsilon^{1/4} \approx 2.2$ GeV, which is indeed bigger than $\Lambda_{\rm QCD}$. We note that the initial energy density, which is estimated above, decreases quickly due to the rapid expansion of the system. The energy density splits into a thermal contribution, which is characterized by a temperature, and one that is related to collective motion. In the local rest frame the thermal energy density is small and its characteristic temperature is comparable to $\Lambda_{\rm QCD}$.

One can also argue using the Color Glass Condensate approach \cite{Gelis:2012ri} that the regime of asymptotic freedom is achieved at the earliest stage of relativistic heavy-ion collisions. 
In this model, the hard scale of saturation $Q_s$ is generated dynamically, and since $Q_s \approx 1-3$ GeV \cite{Gelis:2012ri}, the weak coupling condition $Q_s \gg \Lambda_{\rm QCD}$ is weakly satisfied. 

We conclude therefore that although the smallness of the coupling constant is far from guaranteed, there are good reasons to believe this assumption is justified. 

Our analysis is performed only with an extremely oblate distribution for which the calculations can be done analytically to some extent. In real experiments the anisotropy parameter $\xi$ is certainly finite. However, as shown in Fig.~16 of the review article \cite{Mrowczynski:2016etf}, when $\xi$ is large the growth rate of the unstable modes is rather insensitive to its value. We therefore expect that the magnitudes of $\hat{q}$ obtained using a realistic momentum distribution are not much different from those presented here.   

Another important limitation of our approach is that we computed only the soft contribution to $\hat{q}$ corresponding to momentum transfers smaller than  $k_{\rm max}$. However, as demonstrated in Fig.~\ref{fig-kmax-dependence}, $\hat{q}$ depends logarithmically on the upper cut-off $k_{\rm max}$. Therefore, the hard contribution can be approximately included by shifting $k_{\rm max}$ to the kinematic limit which is roughly the energy of the test parton.  

In this study we have decided to stay away from phenomenology, deferring an analysis of how to obtain quantitative experimental predictions to a future publication. Nevertheless we give here  rough estimates of the parameters which are involved in our calculations, to make our results more useful.

To express the time from Fig.~\ref{qhatversustime} in physical units, one needs a value of the parameter $m$ defined in Eq.~(\ref{mass-qcd}). To get a crude estimate of $m$, we assume that the energy density (\ref{ini-en-den}) corresponds to an ideal gas of gluons in thermal equilibrium 
\be
\varepsilon = \frac{\pi^2(N_c^2 -1)}{15} T^4\,.
\ee
Using $\varepsilon^{1/4} = 2.2$ GeV and $N_c=3$ we have $T = 1.5$ GeV, and substituting this temperature  into the Debye mass formula we obtain 
\be
m = \sqrt{\frac{4 \pi \alpha_s N_c}{3}}\, T = 0.74 \, \varepsilon^{1/4} = 1.7 ~{\rm GeV}\,,
\ee
where we have used $\alpha_s \equiv g^2/4\pi =0.1$.  The time unit is therefore $m^{-1} = 0.12$ fm/c. 

From Fig.~\ref{qhatversustime} one can read off the ratio of $\hat{q}$ in extremely oblate and in equilibrium QGP at the same value of $m$. To get a rough estimate of $\hat{q}_{\rm eq}$ we use the approximate formula \cite{Qin:2015srf}
\be
\hat{q}_{\rm eq} = 2 C_R \alpha_s m^2 \log\Big(\frac{E}{m}\Big) \,,
\ee
where the maximal momentum transfer $k_{\rm max}$ is identified with the energy of the test parton $E$. When the test parton is a gluon ($C_R=3$) with energy $E = 50$ GeV and, as previously, $\alpha_s =0.1$, $m = 1.7$ GeV and $T = 1.5$ GeV, we have $\hat{q}_{\rm eq} = 3.8 \; {\rm GeV^3} = 19 \; {\rm GeV^2/fm}$. 
From Fig.~\ref{qhatversustime} we find for $t = 10/m = 1.2 \; {\rm fm}/c$, for example, $\hat{q} = 5 \,\hat{q}_{\rm eq} = 95 \; {\rm GeV^2/fm}$, if the test parton moves perpendicularly to the beam axis. 

The value of $\hat{q}$ that is required to reproduce the experimentally observed pattern of jet quenching is much smaller, that is $\hat{q} \sim 1-3 \; {\rm GeV^2/fm}$ \cite{Qin:2015srf}. Since the oblate plasma exists only for a short time interval however, the total momentum broadening produced by the effect we have calculated could be rather moderate. 
In any case, our analysis strongly suggests that a significant contribution to the total momentum broadening comes from the brief phase during which the plasma is unstable.

Let us now recapitulate and conclude our study. \rm We have developed a formalism to compute the momentum broadening parameter $\hat q$ in an unstable plasma. Our calculation is formulated as an initial value problem, and we produce a time dependent result which gives the dynamical evolution of $\hat q$. We have calculated the momentum broadening parameter for the case of extremely oblate plasma, which is relevant to the study of heavy ion collisions. We find that $\hat q$ grows exponentially with time, as a result of the interaction of the test parton with the unstable collective modes in the system. At times which are large compared with the inverse fundamental mass scale, the magnitude of $\hat q$ can be much bigger than in equilibrated plasmas. 

We comment that the mere presence of unstable modes is not enough to guarantee this result. The plasma is a complicated system with many constituents interacting in different  ways. This is (in part) reflected in the complicated structure of the integrand of the momentum broadening parameter obtained in this paper. The exponentially growing terms contain oscillating factors, and the integrand has both positive and negative contributions which produce large cancellations. The exponentially growing result which is obtained when all integrals are performed is therefore far from trivial. In the extremely prolate system, for example, the oscillatory behavior of the integrand overwhelms the influence of the unstable modes, and the exponential growth that is found in the oblate system is not seen. 

The key point is that the very large value of $\hat q$ which is produced in our calculation, indicates that the test parton could lose a sizable fraction of its energy in a transient  pre-equilibrium phase of the plasma. The relevance of our result to the phenomenon of jet quenching in relativistic heavy-ion collisions is an issue that must be studied further. Jet quenching is observed at both RHIC and LHC at almost vanishing rapidity in the center of mass of the colliding nuclei. This configuration is shown in Fig.~\ref{fig-config} with the jet momentum transverse to the $z$-axis. Our results indicate that the momentum broadening $\hat q$ is maximal in this configuration. It would be interesting to see if the jet quenching pattern would be changed if the jet axis is tilted in such a way that the near-side jet has a positive (negative) rapidity while the away-side jet has a negative (positive) rapidity. The effect of the unstable modes would then be reduced and the radiative energy loss should be  smaller.

%-----------------------------------------------------------------------
\section*{Acknowledgments}
%-----------------------------------------------------------------------

MEC is supported by the Natural Sciences and Engineering Research Council of Canada. BPS is supported under DOE Contract No. DE-SC0012704 and acknowledges a DOE Office of Science Early Career Award.

\appendix

%%%%%%%%%%%%%%%%%%%%%%%%%%%%%%%%%%%%%%%%%%%%%%%%%%%
\section{Definition of dimensionful scales}
\label{mass-def-section}
%%%%%%%%%%%%%%%%%%%%%%%%%%%%%%%%%%%%%%%%%%%%%%%%%%%

The mass scale defined in Eq. (\ref{mass-def-new}) is used to define our system of units. We scale all dimensional quantities by this parameter, or equivalently we set $m=1$. In this appendix, we explain the physical meaning of this mass scale. We start by discussing equilibrium plasmas, and we furthermore drop the assumption that all particles obey Boltzmann statistics. This is done for the purposes of discussion only. We will obtain an expression for the mass which is familiar from thermal field theory, in order to illustrate its general interpretation. In a QED plasma composed of (massless) electrons and positrons we have (see Eq. (\ref{mass-def-new}))
\bea
m^2 = 2\sum_\sigma q_\sigma^2 \int \frac{d^3 p}{(2\pi)^3} \frac{n_\sigma(\vec p)}{p} 
= 2 e^2 \int \frac{d^3 p}{(2\pi)^3} \frac{n_e(\vec p)+\bar n_e(\vec p)}{p} \,,
\eea
with 
\bea
\label{distro-qed}
n_e(\vec p) = \frac{2}{e^{\beta (p-\mu)}+1}\,,
~~~~~~~~~~\bar n_e(\vec p) = \frac{2}{e^{\beta (p+\mu)}+1}\,,
\eea
where $\beta$ is the inverse temperature, $\mu$ is the quark chemical potential, $p \equiv |\vec p|$ and the factor 2 in the numerator of the Fermi-Dirac distributions is needed to take into account the 2 possible spin states. Direct calculation with $\mu=0$ gives $m^2 = \frac{e^2 T^2}{3} = m_D^2$ where $m_D$ is the Debye mass. 
In a QCD plasma composed of quarks, anti-quarks and gluons we have
\bea
\label{mass-qcd}
m^2 = g^2 \int \frac{d^3 p}{(2\pi)^3} \frac{ n_q(\vec p)+\bar n_q(\vec p)+n_g(\vec p)}{p} 
\eea
with 
\bea
\label{distro-qcd}
n_q (\vec p) = \frac{2 N_f}{e^{\beta (p-\mu)}+1}\,,~~~~~~
\bar n_q(\vec p) = \frac{2 N_f}{e^{\beta (p+\mu)}+1}\,,~~~~~~
n_g(\vec p) = \frac{4 N_c}{e^{\beta p}-1}\,.
\eea
The factors in Eqs. (\ref{mass-qcd}, \ref{distro-qcd}) can be verified by calculating the QCD Debye mass: with $\mu=0$ the result is $m^2 = g^2 T^2 \big(\frac{N_c}{3}  +\frac{N_f}{6} \big) = m_D^2$. 

The factor $d_{\rm scale}$ defined in Eq.~(\ref{dscale-def}) has dimension mass cubed and contains all of the dimensions in the result for $\hat q$ (the expression in the right side of Eq.~(\ref{qhat5}) divided by $d_{\rm scale}$ is dimensionless). The parameter $d_{\rm scale}$ is related to the mass parameter (\ref{mass-def-new}) that we have used to define our units and a scale that characterizes the width of the distribution of transverse momenta. We assume that all species of plasma particles have an oblate distribution as in Eq. (\ref{ext-oblate}) and therefore the formula (\ref{dscale-def}) can be rewritten
\bea
d_{\rm scale} = \frac{1}{(2\pi)^2} \sum_\sigma q_\sigma^2
\int_0^\infty dp_\perp\, p_\perp h_\sigma(p_\perp) \,,
\eea
and the equation that defines the mass parameter (\ref{mass-def-new}) has the form
\bea
\label{mass-def2}
m^2 = \frac{2}{(2\pi)^2} \sum_\sigma q_\sigma^2 
\int_0^\infty dp_\perp h_\sigma(p_\perp) \,.
\eea
Defining the mean transverse momentum as
\bea
\langle p_\perp\rangle \equiv \frac{\sum_\sigma\int_0^\infty dp_\perp\, p_\perp\,h_\sigma(p_\perp)}
{\sum_\sigma\int_0^\infty dp_\perp\,h_\sigma(p_\perp)} ,
\eea
we obtain 
\bea
d_{\rm scale} = \frac{1}{2}\,m^2  \langle p_\perp\rangle ,
\eea
which shows the physical meaning of the parameter $d_{\rm scale}$. For example, taking the Boltzmann limit of the distributions in Eq. (\ref{distro-qed}) or (\ref{distro-qcd}) would produce $d_{\rm scale} = T\,m_D^2$.

%%%%%%%%%%%%%%%%%%%%%%%%%%%%%%%%%%%%%%%%%%%%%%%%%%%
\section{Removing zero poles}
\label{zero-poles-section}
%%%%%%%%%%%%%%%%%%%%%%%%%%%%%%%%%%%%%%%%%%%%%%%%%%%

The method used in Sec.~\ref{correlator-section} to solve self-consistently the set of Vlasov and Maxwell equations can be summarized as follows: we perform a one sided Fourier transform to rewrite our equations in momentum space, solve the resulting set of equations algebraically, and then perform the inverse transform to rewrite the results in position space. 
In this appendix we discuss a subtle point associated with this procedure. 

We consider the homogeneous Maxwell equation which relates the electric and magnetic fields. This equation is the second part of Eq.~(\ref{max-coord-homo}), which becomes in momentum space Eq.~(\ref{max-k-2}). 
The Fourier transform on the spatial coordinates is not involved in this issue and can be removed from this discussion. We define the functions
\bea
\label{bif-not}
b(t) \equiv B^i(t,\vec k)\,,
~~~~~~~~~~ 
e(t) \equiv -i\epsilon^{ijl}k^j E^l(t,\vec k)
\eea
and rewrite Faraday's law in the simple form
\be
\label{bdif}
\frac{db(t)}{dt} = e(t)\,.
\ee
The equation is supplemented with the initial condition $b(t=0) = b_0$.

We can find the solution of Eq.~(\ref{bdif}) using the procedure outlined in the first paragraph of this section. Taking the one-sided Fourier transform of Eq.~(\ref{bdif}) gives
\bea
\label{bdif2}
 -b_0 -i\omega \, b(\omega) = e(\omega) \,.
\eea
The l.h.s. of Eq.~(\ref{bdif2}) has been found performing the integration by parts as
\be
\int_0^\infty dt\; e^{i\omega t} \frac{db(t)}{dt} =  
e^{i\omega t}b(t) \bigg|^\infty_0 - i\omega \int_0^\infty dt\;e^{i\omega t}b(t) =
-b_0 -i\omega \, b(\omega) \,,
\ee
where the function $b(t)$ is assumed to vanish when $t \to \infty$. 

The solution of the algebraic equation (\ref{bdif2}) is
\bea
\label{res-ft-pre}
b(\omega) = -\frac{b_0+e(\omega)}{i\omega}
\eea
and taking the inverse transform we obtain
\bea
\label{res-ft}
b(t) &=& -\int^{\infty+i \sigma}_{-\infty+i \sigma}\frac{d\omega}{2\pi i}e^{-i\omega t}\frac{b_0}{\omega}-\int^{\infty+i \sigma}_{-\infty+i \sigma}\frac{d\omega}{2\pi i}\, e^{-i\omega t} \,\frac{e(\omega)}{\omega}
\nonumber \\[4mm]
&=& b_0-\int^{\infty+i \sigma}_{-\infty+i \sigma}\frac{d\omega}{2\pi i}\,e^{-i\omega t}\,\frac{e(\omega)}{\omega} \,,
\eea
where the result for the first integral is obtained by noticing that the only contribution comes from the pole at $\omega=0$. 
Equation (\ref{res-ft-pre}) is exactly the same as Eq.~(\ref{max-k-2}), merely rewritten using the simplified notation defined in Eq.~(\ref{bif-not}). 

Notice that the integral on the right side of Eq.~(\ref{res-ft}) must be zero at $t=0$, in order to produce $b(0)=b_0$. This means that physical solutions for the electric field must satisfy 
\bea
\label{cond-e-fun}
\int^{\infty+i \sigma}_{-\infty+i \sigma}\frac{d\omega}{2\pi i}\,\frac{e(\omega)}{\omega} = 0 \,,
\eea
and if this is not true then the initial condition $b(0) = b_0$ is incompatible with Eq.~(\ref{bdif}). It is not difficult to invent such a situation. For example, if $e(t)$ diverges as $t^{-1}$ when $t \to 0$,  the initial value $b(0)$ cannot be finite. Using the condition (\ref{cond-e-fun}), the solution (\ref{res-ft}) could be rewritten in an equivalent form
\bea
\label{res-ft2}
b(t) = b_0-\int^{\infty+i \sigma}_{-\infty+i \sigma}\frac{d\omega}{2\pi i}\,\big[e^{-i\omega t}-1\big]\,\frac{e(\omega)}{\omega}\,.
\eea

The result (\ref{res-ft2}) can be obtained in a different way by solving the trivial differential equation (\ref{bdif}) as
\bea
b(t) = b_0+\int^t_0 dt^\prime\;e(t^\prime)\,,
\eea
and expressing the function $e(t)$ through its Fourier transform which gives
\bea
b(t) = b_0 + \int^t_0 dt^\prime \bigg[\int^{\infty+i \sigma}_{-\infty+i \sigma} \frac{d\omega}{2\pi} e^{-i\omega t^\prime} e(\omega)\bigg]\,.
\eea
Switching the order of integrations over $t'$ and $\omega$ and performing the time integral explicitly, we again obtain the formula (\ref{res-ft2}). 

Let us compare the two forms (\ref{res-ft})  and (\ref{res-ft2}) of the solution of the differential equation (\ref{bif-not}). One observes that if the function $e(\omega)$ is regular at $\omega = 0$, the integrand in the formula (\ref{res-ft}) has a pole $\omega =0$, but the integrand in the formula (\ref{res-ft2}) is regular. In general, when  $e^{-i\omega t}$ is replaced by  $(e^{-i\omega t}-1)$, the order of the pole of the integrand at $\omega =0$ is reduced by one. 

We have found that the terms in the integrand in Eq.~(\ref{qhat4}) which come from the correlators involving the magnetic field diverge at $\omega =0$. We treat this as a signal that our initial conditions are incompatible with Faraday's law and we use the freedom to choose the solution in the form (\ref{res-ft}) or (\ref{res-ft2}) to obtain a finite expression.  Therefore, the exponential factors $e^{-i\omega t}$ which enter the functions ${\cal I}_{XX}(t)$ and ${\cal I}_{XY}(t)$ with $\{X,Y\}\in\{E,B\}$ in Eqs.~(\ref{MBexponentialI}, \ref{EB-exp}, \ref{BE-exp}, \ref{BB-exp}), are changed to $(e^{-i\omega t}-1)$ when magnetic fields are present in the correlators.

%%%%%%%%%%%%%%%%%%%%%%%%%%%%%%%%%%%%%%%%%%%%%%%%%%%
\section {Equilibrium limit}
\label{equib-appendix}
%%%%%%%%%%%%%%%%%%%%%%%%%%%%%%%%%%%%%%%%%%%%%%%%%%%

In this Appendix we derive the momentum broadening parameter $\hat q$ for equilibrium plasmas. The result predicted by the classical Langevin approach was obtained previously in \cite{Majumder:2009cf} directly using the equilibrium field correlators derived in \cite{Mrowczynski:2008ae}. The aim of this Appendix is to verify that our formula (\ref{qhat4}), which is used to compute $\hat q$  for anisotropic plasmas, reduces to the correct expression in the equilibrium limit. 

The factors ${\cal C}_{EE}$, ${\cal C}_{EB}$, ${\cal C}_{BE}$ and ${\cal C}_{BB}$, which enter the formula (\ref{qhat4}), are calculated using the procedure described in Sec.~\ref{integrand-section}. When calculating the integrand, the difference between the equilibrium and oblate integrands is that the propagator $\Delta^{ij}(\omega,\vec k)$, which enters through equations (\ref{elec3}, \ref{magnetic3}), is given by equation (\ref{equib-prop}) in equilibrium and (\ref{anio-prop}) in anisotropic plasma. 

The integral in Eq. (\ref{MBexponentialI}) can be calculated analytically (the results for the integrals in Eqs.~(\ref{EB-exp} - \ref{BB-exp}) will not be needed - this is explained below). Direct integration produces
\bea
\label{IEE-equib}
{\cal I}_{EE}(t) = -\frac{i}{k}\bigg[\frac{e^{-i k t(\hat\omega_2+ \vec u\cdot\hat k)}}{\hat\omega_1 - \vec u\cdot\hat k} + \frac{e^{-i k t(\hat\omega_1 - \vec u\cdot\hat k)}}{\hat\omega_2 + \vec u\cdot\hat k} + \frac{e^{-i k t(\hat\omega_1+\hat\omega_2)}(\hat\omega_1 + \hat\omega_2)}{(\hat\omega_1 - \vec u\cdot\hat k)(\hat\omega_2 + \vec u\cdot\hat k)}  \bigg]\,.
\eea

The next step is to perform the frequency integrals. All collective modes are damped in equilibrium and therefore give contributions to $\hat q$ which exponentially decay in time. Consequently we include only the contributions from the Landau poles at $\omega_1 = \vec k \cdot \vec v$ and $\omega_2 = -\vec k \cdot \vec v$. After substituting the expressions for the Landau poles, the factor in Eq.~(\ref{IEE-equib}) takes the simple form
\bea
{\cal I}_{EE}(t)\bigg|_{\substack{\hat \omega_1=\hat k\cdot \vec v \\ \hat \omega_2=-\hat k\cdot \vec v}} 
= \frac{2\sin\big(t \;\vec k \cdot(\vec v-\vec u)\big)}{\vec k \cdot(\vec v-\vec u)}\,.
\eea
In the equilibrium calculation we are dealing with a static medium and therefore we take the long time limit to eliminate short time switching-on effects. It is easy to show that 
\bea
\label{deltafunction-res}
\lim_{t\to\infty} {\cal I}_{EE}(t) = 2\pi\delta\big(\vec k \cdot(\vec u-\vec v)\big)\,.
\eea
In fact, we can immediately see how this delta function arises. The integral in Eq. (\ref{MBexponentialI}) gives zero in the long time limit unless the phase of the exponentials is zero, and when the frequencies take the values $\hat \omega_1=\hat k\cdot \vec v$ and $\hat \omega_2=-\hat k\cdot \vec v$ this means we require $\vec k\cdot\vec u = \vec k\cdot\vec v$. In the same way we see that the extra terms that are introduced by the $-1$ terms in Eqs. (\ref{EB-exp} - \ref{BB-exp}) give zero in the long time limit, and likewise any potential contribution from a zero pole would give zero at long times. The conclusion is that in the equilibrium calculation the square bracket in (\ref{qhat4}) becomes 
\bea
\label{Ceq-def}
\lim_{t\to\infty} {\cal I}_{EE}(t) \;\big[{\cal C}_{EE} +  {\cal C}_{EB} 
+ {\cal C}_{BE} +  {\cal C}_{BB}\big]  = 2\pi \delta\big(\vec k\cdot(\vec u-\vec v)\big)\;{\cal C}_{\rm eq}
\eea
where we have defined 
\bea
{\cal C}_{\rm eq} \equiv {\cal C}_{EE} + {\cal C}_{EB} 
+  {\cal C}_{BE} + {\cal C}_{BB}\,.
\eea 
The factor ${\cal C}_{\rm eq}$ is calculated using our {\it Mathematica} program. 

We use the definitions (\ref{dscale-def}) and (\ref{Ceq-def}) to rewrite the formula (\ref{qhat4}). We also introduce unity in the form $\int d\omega \,\delta(\omega-\vec v\cdot\vec k)$ and replace all factors $\vec v\cdot\vec k$ with $\omega$. This produces the expression
\be
\label{equib1}
\hat q_{\rm eq} = (2\pi)^2 \, e^2 \int \frac{d\omega}{2\pi}  \int\frac{d^3 k}{(2\pi)^3} \;\delta(\omega-\vec u\cdot\vec k) \, J\,,
\ee
with
\be
\label{J-def}
J \equiv \sum_\sigma q_\sigma^2\int \frac{d^3p}{(2\pi)^3} \,
\delta(\omega-\vec k\cdot\vec v) \,k^2 \, {\cal C}_{\rm eq} \, n_\sigma(p)\,.
\ee

In order to do the integral over $\vec p$ in Eq.~(\ref{J-def}) we will make use of the fact that the equilibrium distribution depends only on the magnitude of $\vec p$. The first step is to rewrite the result for ${\cal C}_{\rm eq}$ as
\bea
\label{Ceq}
&& {\cal C}_{\rm eq} = T^{ij}(M^{ij}_{EE}+ M^{ij}_{BB})+S^{ij}(-M^{ij}_{EB}+M^{ij}_{BE}) \,,
\eea
where we have defined
\be
\label{parton-projectors}
T^{ij} \equiv \delta^{ij}-u^i u^j 
~~~~~~~~ \text{and} ~~~~~~~~
S^{ij} \equiv \epsilon^{ijm}u^m \,.
\ee
The tensors $M_{EE}$, $M_{BB}$, $M_{EB}$ and $M_{BE}$, which are produced by our program, are
\begin{subequations}
\label{iso-corr}
\bea
M^{ij}_{EE}  &=& \hat\omega^4 \hat k^i \hat k^j \Delta_L(\omega)\Delta_L(-\omega)
+\hat\omega^3 (v^i-\hat\omega\hat k^i) \hat k^j  \Delta_T(\omega) \Delta_L(-\omega) 
\\ \nn
&& + \, \hat\omega^3 (v^j-\hat\omega\hat k^j) \hat k^i \Delta_L(\omega)\Delta_T(-\omega) 
+\hat\omega^2 (\hat\omega \hat k^i-v^i)(\hat\omega \hat k^j-v^j) \Delta_T(\omega)\Delta_T(-\omega) \,, 
\\[4mm]
M^{ij}_{BB} &=& \Delta_T(\omega)\Delta_T(-\omega)\big[ \delta_{ij}\big(1-(\vec v\cdot\hat k)^2\big) -\hat k^i\hat k^j -\vec v^i\vec v^j + \vec v\cdot\hat k(v^i \hat k^j +\hat k^i v^j ) \big] \,,
\\[4mm]
M^{ij}_{EB} &=& (\vec v\cdot\hat k)^2 \hat k^i (\hat k \times \vec v)^j \;\Delta_L(\omega)\Delta_T(-\omega)
\\ \nn
&& ~~~~~~~~~~~~~~~~~~~~~~~~~~~~~~~~
+ \, \vec v\cdot\hat k (v^i - \vec v\cdot\hat k \hat k^i)(\hat k \times \vec v)^j \; \Delta_T(\omega)\Delta_T(-\omega) \,, 
\\[4mm]
M^{ij}_{BE} &=& (\vec v\cdot\hat k)^2 \hat k^j (\hat k \times \vec v)^i \; \Delta_T(\omega)\Delta_L(-\omega) \,,
\\ \nn
&& ~~~~~~~~~~~~~~~~~~~~~~~~~~~~~~~~
+ \, \vec v\cdot\hat k (v^j - \vec v\cdot\hat k \hat k^j)(\hat k \times \vec v)^i \; \Delta_T(\omega)\Delta_T(-\omega)\,.
\eea
\end{subequations}
The argument $\vec k$ of the propagators is suppressed in Eqs. (\ref{iso-corr}) and $\hat\omega \equiv \omega/k$ with $k\equiv |\vec k|$ and $\hat k^i \equiv k^i/k$. Using an obvious notation we divide the integral $J$ into four pieces which we call $J_{EE}$, $J_{BB}$, $J_{EB}$ and $J_{BE}$. These four contributions to $J$ can be written 
\begin{subequations}
\label{JXX}
\bea
&& J_{XX} = T^{ij}\,I^{ij}_{XX}\,, ~~~~~~~~~
I^{ij}_{XX} = k^2\, \sum_\sigma q_\sigma^2\int\frac{d^3p}{(2\pi)^3}\; 
\delta(\omega-\vec k\cdot\vec v)  n_\sigma(p) M^{ij}_{XX} \,,
\\
&& J_{XY} = C^{ij}\,I^{ij}_{XY} \,, ~~~~~~~~~
I^{ij}_{XY} = k^2\, \sum_\sigma q_\sigma^2\int\frac{d^3p}{(2\pi)^3}\; 
\delta(\omega-\vec k\cdot\vec v) n_\sigma(p) M^{ij}_{XY} \,,
\eea
\end{subequations}
where $J_{XX}$ means $J_{EE}$ or $J_{BB}$ and $J_{XY}$ means $J_{EB}$ or $J_{BE}$.
We define
\bea
\label{Ihtl-def}
&& I_{\rm thermal} \equiv \sum_\sigma q_\sigma^2 \int \frac{d^3p}{(2\pi)^3} \delta(\omega-\vec k\cdot\vec v)  n_\sigma(p)\,.
\eea

We want to factor the pieces $M^{ij}_{XX}$ and $M^{ij}_{XY}$ from the integrals in Eq.~(\ref{JXX}) and extract $I_{\rm thermal}$. The problem is that they contain factors that depend explicitly on the velocity $\vec v$ and therefore cannot be pulled out of the integral over $\vec p$. We can show however that, due to plasma isotropy, these factors disappear when the $\vec p$ integral is done.  
The proof is as follows. In an isotropic plasma, a symmetric correlator of the form $I^{ij}_{XX}$ must be a linear combination of the two projectors $A^{ij}$ and $B^{ij}$ in equation (\ref{AB-def}). Thus we have
\bea
\label{contrXX}
I^{ij}_{XX}=\frac{1}{2}A^{ij}(A^{lm}I^{lm}_{XX}) + B^{ij}(B^{lm}I^{lm}_{XX})\,.
\eea
Similarly, the antisymmetric combination $I^{ij}_{XY}-I^{ij}_{YX}$ must be proportional to the tensor
\bea
F^{ij}=\epsilon^{ijl}\frac{k^l}{k} \,,
\eea
and therefore we write
\bea
\label{contrXY}
I^{ij}_{XY}-I^{ij}_{YX}=\frac{1}{2}F^{ij}\big[F^{lm}(I^{lm}_{XY}-I^{lm}_{YX})\big]\,.
\eea
The integral $I_{\rm thermal}$ can be factored from the contacted expressions $A^{lm}I^{lm}_{XX}$, $B^{lm}I^{lm}_{XX}$ and $F^{lm}(I^{lm}_{XY}-I^{lm}_{YX})$. The results are
\begin{subequations}
\label{contrs}
\bea
A^{lm}I^{lm}_{EE} &=& \frac{1}{2} I_{\rm thermal} (\vec u\cdot\hat k)^2 \big(1-(\vec u\cdot\hat k)^2\big)  \Delta_T(\omega,\vec k)\Delta_T(-\omega,\vec k) \,,
\\[2mm]
B^{lm}I^{lm}_{EE} &=& I_{\rm thermal} (\vec u\cdot\hat k)^4 \Delta_L(\omega,\vec k)\Delta_L(-\omega,\vec k) \,,
\\[2mm]
A^{lm}I^{lm}_{BB} &=& \frac{1}{2} I_{\rm thermal} \big(1-(\vec u\cdot\hat k)^2\big) \Delta_T(\omega,\vec k)\Delta_T(-\omega,\vec k) \,, 
\\[2mm]
B^{lm}I^{lm}_{BB}  &=& 0 \,,
 \\[2mm]
F^{lm}(I^{lm}_{EB} -I^{lm}_{BE}) &=&  I_{\rm thermal} (\vec u\cdot\hat k) \big(1-(\vec u\cdot\hat k)^2\big) \Delta_T(\omega,\vec k)\Delta_T(-\omega,\vec k) \,. 
\eea
\end{subequations}
Combining equations (\ref{parton-projectors}, \ref{JXX}, \ref{contrXX}, \ref{contrXY}, \ref{contrs}), it is straightforward to show that the expression (\ref{Ceq}) can be written as
\be
J = k^2 \; I_{\rm thermal}\;{\cal C}_{\rm eq}\,
\ee
with
\be \nn
{\cal C}_{\rm eq} \equiv
\frac{1}{2} \big(1-(\vec u\cdot\hat k)^2\big)^3 \Delta_T(\omega,\vec k)\Delta_T(-\omega,\vec k)
+
(\vec u\cdot\hat k)^4 \big(1-(\vec u\cdot\hat k)^2\big) \Delta_L(\omega,\vec k)\Delta_L(-\omega,\vec k)\,.
\ee

Using the delta function to do the frequency integral and defining $k_z = \vec u \cdot \vec k$ (the axis $z$ is chosen along the velocity of the test parton), the result for $\hat q$ in Eq.~(\ref{equib1}) takes the form
\bea
\label{qhat-eq1}
\hat q_{\rm eq} &=& 2\pi \, e^2  \int\frac{d^3 k}{(2\pi)^3} \;k^2\; I_{\rm thermal}\Big|_{\omega=k_z}
\\[2mm]\nn
&\times& \Big[\hat k_z^4\big(1-\hat k_z^2\big) \Delta_L(k_z,\vec k)\Delta_L(-k_z,\vec k) +\frac{1}{2}\big(1-\hat k_z^3\big)\Delta_T(k_z,\vec k)\Delta_T(-k_z,\vec k) \Big]\,.
\eea
Introducing the scale parameter (\ref{dscale-def}) and referring to the definition (\ref{Ihtl-def}), this result can be rewritten as
\bea
\label{qhat-eqnum}
\hat q_{\rm eq} &=& 2\pi \, e^2  d_{\rm dscale} \int\frac{d^3 k}{(2\pi)^3}\;\frac{k}{2} 
\\ [2mm] \nn
&\times& \Big[\hat k_z^4\big(1-\hat k_z^2\big) \Delta_L(k_z,\vec k)\Delta_L(-k_z,\vec k) +\frac{1}{2}\big(1-\hat k_z^3\big)\Delta_T(k_z,\vec k)\Delta_T(-k_z,\vec k) \Big]\,.
\eea
The integral can be calculated numerically and for $k_{\rm max}= 2m$ gives $\hat q =  0.11 \, e^2 d_{\rm scale}$ which is a reference point for our time dependent results on oblate plasma which are discussed in Sec.~\ref{results-section}.

In order to compare with the result derived in \cite{Mrowczynski:2008ae}, we rewrite Eq.~(\ref{qhat-eq1}). First we note that equations (\ref{PI}, \ref{pi-decomposition-iso}) can be used to rewrite the quantity (\ref{Ihtl-def}) as
\bea
\label{Ihtl-def-2}
I_{\rm thermal} = -\frac{2T\;{\rm Im}\beta(\omega,\vec k)}{\pi\hat\omega^3 k} = \frac{4T \;{\rm Im}\alpha(\omega,\vec k)}{\pi \hat\omega(\hat\omega^2-1)k}\,.
\eea
Using the formulas (\ref{Delta-T-L}, \ref{Ihtl-def-2}) and the symmetry relations
\begin{subequations}
\bea
{\rm Re}\alpha(k_z,\vec k) = {\rm Re}\alpha(-k_z,\vec k)\,,~~~~~~ 
{\rm Im}\alpha(k_z,\vec k) = -\,{\rm Im}\alpha(-k_z,\vec k) \,,
\\[2mm]
{\rm Re}\beta(k_z,\vec k) = {\rm Re}\beta(-k_z,\vec k)\,,~~~~~~ 
{\rm Im}\beta(k_z,\vec k) = -\,{\rm Im}\beta(-k_z,\vec k) \,,
\eea
\end{subequations}
Eq.~(\ref{Ihtl-def-2}) gives
\begin{subequations}
\label{Ihtl-def-3}
\bea
&& I_{\rm thermal} \;\Delta_T(\omega,\vec k)\Delta_T(-\omega,\vec k) 
= \frac{2T}{\pi k\hat\omega(\hat\omega^2-1)}\;{\rm Im}\Delta_T(\omega,\vec k)\,,
\\[2mm]
&& I_{\rm thermal} \;\Delta_L(\omega,\vec k)\Delta_L(-\omega,\vec k) 
= -\frac{T}{\pi k\hat\omega^3}\;{\rm Im}\Delta_L(\omega,\vec k)\,,
\eea
\end{subequations}
and substituting (\ref{Ihtl-def-3}) into (\ref{qhat-eq1}) we obtain:
\bea
\label{qhat-eq2}
\hat q_{\rm eq} = -2 T e^2\int\frac{d^3k}{(2\pi)^3}\frac{\big(k^2-k_z^2\big)}{k_z}\big[\big(1-\hat k_z^2\big) {\rm Im}\Delta_T(k_z,\vec k) + \hat k_z^2 \Delta_L(k_z,\vec k)\big]\,.
\eea

The final step is to rewrite the result (\ref{qhat-eq2}) in terms of the dielectric tensor $\varepsilon_{ij}(\omega,\vec k)$ which is related to the polarization tensor as
\bea
\label{dielec-def}
\varepsilon^{ij}(\omega,\vec k) = \delta^{ij}-\frac{\Pi^{ij}(\omega,\vec k)}{\omega^2}\,.
\eea
Transverse and longitudinal components of the dielectric tensor are defined as usual with the projection operators in equation (\ref{AB-def})
\bea
\label{eTeL-def}
\varepsilon_T(\omega,\vec k) = \frac{1}{2}A^{ij}\varepsilon^{ji}(\omega,\vec k)\,,
~~~~~~~~~~~
\varepsilon_L(\omega,\vec k) = B^{ij}\varepsilon^{ji}(\omega,\vec k)\,.
\eea
Equations (\ref{sd-def}, \ref{equib-prop}, \ref{dielec-def}, \ref{eTeL-def}) give
\bea
&& \Delta^{-1}_T(\omega,\vec k) = \omega^2 \varepsilon_T(\omega,\vec k) -k^2\,,
~~~~~~
\Delta^{-1}_L(\omega,\vec k) = \omega^2 \varepsilon_L(\omega,\vec k)\,,
\eea
and Eq. (\ref{qhat-eq2}) takes the form
\bea
\label{qhat-stan}
&& \hat q_{\rm eq}  = 
2 T e^2 \int \frac{d^3k}{(2\pi)^3} \: 
\frac{(k^2-k_z^2)}{k_z k^2}   \bigg[ \frac{{\rm Im} \varepsilon_L(k_z,\vec k)}
{|\varepsilon_L(k_z,\vec k)|^2} 
+ \frac{ k_z^2 (k^2-k_z^2) \: {\rm Im} \varepsilon_T(k_z,\vec k)}
{| k_z^2 \varepsilon_T(k_z,\vec k) - k^2|^2} \bigg] \,.
\eea
As explained in Sec.~\ref{QGP-section}, the result is converted to the corresponding QCD expression by multiplying by the color factor $C_R$ and replacing the QED coupling constant $e$ with the QCD coupling $g$. After making these replacements, Eq. (\ref{qhat-stan}) agrees with equation (27) given in \cite{Mrowczynski:2008ae}. 

%%%%%%%%%%%%%%%%%%%%%%%%%%%%%%%%%%%%%%%%%%%%%%%%%%%%%%%%
\section{Propagator and polarization tensor}
\label{prop-appendix}
%%%%%%%%%%%%%%%%%%%%%%%%%%%%%%%%%%%%%%%%%%%%%%%%%%%%%%%

In this Appendix we discuss the gauge boson propagator and polarization tensor in anisotropic plasma. We start with a brief discussion of isotropic plasma where the polarization tensor (and any symmetric tensor that depends on the wave vector $\vec k$), can be decomposed as
\bea
\label{pi-decomposition-iso}
\Pi^{ij}(\omega,\vec k) = \alpha(\omega,\vec k) A^{ij}+\beta(\omega,\vec k) B^{ij}\,,
\eea
where the two projection operators are defined as
\bea
\label{AB-def}
A^{ij} \equiv \delta^{ij} - \frac{k^i k^j}{k^2}\,,~~~~~~
B^{ij} \equiv \frac{k^i k^j}{k^2} \,.
\eea
The inverse propagator in temporal axial gauge (which we use in Eq. (\ref{full-prop})) equals
\bea
\label{sd-def}
(\Delta^{-1})^{ij}(\omega,\vec k) && = (\Delta^{-1}_{\rm bare})^{ij}(\omega,\vec k) - \Pi^{ij}(\omega,\vec k)  
\\[1mm]
&& =  \big(\omega^2-k^2-\alpha(\omega,\vec k)\big) A^{ij}
+ \big(\omega^2-\beta(\omega,\vec k)\big) B^{ij} \,,\nonumber
\eea
where the inverse bare propagator is given by Eq.~(\ref{bare-prop-inv}). Inverting the tensor (\ref{sd-def}) gives the well known result
\be
\label{equib-prop}
\Delta^{ij}(\omega,\vec k) = \Delta_T(\omega,\vec k) \, A^{ij} +\Delta_L(\omega,\vec k)\,  B^{ij} \,,
\ee
with 
\be
\label{Delta-T-L}
\Delta_T^{-1}(\omega,\vec k) \equiv \omega^2-k^2-\alpha(\omega,\vec k)  \,,
~~~~~~~~~~~~
 \Delta_L^{-1}(\omega,\vec k) \equiv \omega^2-\beta(\omega,\vec k) \,.
\ee
The dispersion equations of transverse longitudinal plasmons are
\be
\Delta_T^{-1}(\omega,\vec k)  = 0 \,, ~~~~~~~~~~~~~~
\Delta_L^{-1}(\omega,\vec k)  = 0 \, .
\ee

In a plasma with a momentum distribution obtained from the isotropic one by stretching or squeezing along the (unit) vector $\vec n$, we need to introduce two additional operators which are defined as
\bea
\label{CD-def}
C^{ij} = \frac{n_T^i n_T^j}{n_T^2}\,,~~~~~~~~~~~~
D^{ij} = k^i n_T^j + k^j n_T^i\,,
\eea
where $n_T^i \equiv A^{ij} \, n^j$. The four operators (\ref{AB-def}, \ref{CD-def}) form a complete basis
(but $D$ does not satisfy $D^2 = D$ and therefore should not be called a projection operator).

The polarization tensor is decomposed as
\bea
\label{pi-decomposition-anio}
\Pi^{ij}(\omega,\vec k) = \alpha(\omega,\vec k) A^{ij}+\beta(\omega,\vec k) B^{ij} + \gamma(\omega,\vec k) C^{ij} + \delta(\omega,\vec k) D^{ij} \,,
\eea
and the inverse propagator and its inversion are
\bea
&& (\Delta^{-1})^{ij} = (\omega^2-k^2-\alpha ) A^{ij}
+ (\omega^2-\beta) B^{ij} - \gamma C^{ij} - \delta D^{ij} \,,
\\[2mm]
\label{anio-prop}
&& \Delta^{ij}= \Delta_A \, (A^{ij}-C^{ij})
+  \Delta_G \Big[(\omega^2 - k^2 -\alpha - \gamma )B^{ij} 
+ \big(\omega^2-\beta \big) C^{ij} + \delta D^{ij} \Big]\,,
\eea
where the arguments $\omega,\vec k$ are suppressed and 
\bea
\label{Delta-A}
&& \Delta_A^{-1}(\omega,\vec k)  \equiv  \omega^2 - k^2 - \alpha(\omega,\vec k)  \,, \\[2mm]
\label{Delta-G}
&& \Delta^{-1}_G(\omega,\vec k) \equiv \big(\omega^2 - \beta(\omega,\vec k)\big)
\big[\omega^2 - k^2 - \alpha(\omega,\vec k) - \gamma(\omega,\vec k)\big]
- k  n_T^2 \delta^2(\omega,\vec k)\,.
\eea
The plasmon dispersion equations are
\be
\label{dis-eqs-aniso}
\Delta_A^{-1}(\omega,\vec k)  = 0 \,, ~~~~~~~~~~~~~~
\Delta_G^{-1}(\omega,\vec k)  = 0 \, .
\ee
The complete plasmon spectrum given by Eqs.~(\ref{dis-eqs-aniso}) is analyzed in detail for all possible degrees of one dimensional deformation of an isotropic momentum distribution in our extensive study \cite{Carrington:2014bla}.

%%%%%%%%%%%%%%%%%%%%%%%%%%%%%%%%%%%%%%%%%%%%%%%%%%%
\section{Integrals over the momenta of plasma particles}
\label{pints-appendix}
%%%%%%%%%%%%%%%%%%%%%%%%%%%%%%%%%%%%%%%%%%%%%%%%%%%

We explain here how to do the integral over $\vec p$ in equation (\ref{qhat4}). The method is the same for every term in ${\cal C}_{XY}$ with $\{X,Y\}\in\{E,B\}$. Since the plasma constituents are assumed massless, the angular integrals factor from the integral over $p \equiv |\vec p|$. The integrand depends only on the azimuthal angle $\varphi$, and all $\varphi$ dependence  comes from factors $\vec u \cdot \vec v$ and $\hat k \cdot \vec v$. We define a generic $\varphi$ integral
\be
\label{generic-phi1}
I^{k\,l}_{m\, n_1 \, n_2}(\hat\omega_1,\hat\omega_2,\vec k) 
\equiv \int_0^{2\pi} d\varphi ~ \frac{[\sin\varphi]^k~[\cos\varphi]^l\,[C(\varphi)]^m}{ [D_-(\hat\omega_1,\varphi)]^{n_1} D_+(\hat\omega_2,\varphi)]^{n_2}}\,,
\ee
where the numbers $k,l,m,n_1,n_2$ are integer and 
\bea
\label{generic-phi1-b}
&& C(\varphi) \equiv \frac{1}{1+m_{\rm min}^2-(\vec v\cdot \hat k)^2} = \frac{1}{1+m_{\rm min}^2-(1-x^2)\cos^2\varphi}\,,\\[4mm] 
&& D_\pm(\hat\omega,\varphi) \equiv (\hat\omega + i\epsilon \pm \vec v\cdot\hat k) = (\hat\omega + i\epsilon \pm \sqrt{1-x^2}\cos\varphi)\,,
\eea
$\epsilon$ is an infinitesimally small real positive number and $x\equiv \cos\theta$ with $\theta$ being the angle between the vector $\vec k$ and axis $z$.  As discussed in Sec.~\ref{integral-section}, the parameter $m_{\rm min}$ is introduced in the definition of $C(\varphi)$ to regulate the divergence in the $\varphi$ integral when $x=0$. 

All of the integrals of the form (\ref{generic-phi1}) can be done analytically, but it is more efficient to rearrange them into a simpler form. Difficulties are caused by the factors $D_-(\hat\omega_1,\varphi)$ and $D_+(\hat\omega_2,\varphi)$ which contain zeros. In many terms these denominators can be removed using a simple trick. First we rewrite
\bea
\label{pfrule}
\frac{1}{D_-(\hat\omega_1,\varphi)D_+(\hat\omega_2,\varphi)} = \frac{1}{\hat\omega_1+\hat\omega_2}\bigg[\frac{1}{D_-(\hat\omega_1,\varphi)} + \frac{1}{D_+(\hat\omega_2,\varphi)} \bigg]\,.
\eea
Second we remove a factor $\cos\varphi$ in the numerator using
\begin{subequations}
\label{num-example}
\begin{align}
\frac{\cos\varphi}{D_+(\hat\omega_2,\varphi)} = \frac{1}{\sqrt{1-x^2}}\bigg[1-\frac{\hat\omega_2}{D_+(\hat\omega_2,\varphi)}\bigg] \,,\\[4mm] 
 \frac{\cos\varphi}{D_-(\hat\omega_1,\varphi)} = \frac{1}{\sqrt{1-x^2}}\bigg[\frac{\hat\omega_1}{D_-(\hat\omega_1,\varphi)}-1\bigg]\,.
\end{align}
\end{subequations}
We proceed by systematically partial fractioning and removing $\cos\varphi$ factors in the numerator using the formulas (\ref{pfrule}, \ref{num-example}). Most of the remaining terms have no poles and can be easily evaluated. 

Notice however that the expression produced by partial fractioning (\ref{pfrule}) will cause problems at the next step, which will be to perform the $\omega_1$ and $\omega_2$ integrals using contour integration. Each integral will get contributions from each of the poles listed in Table \ref{mode-table}. However, when we take the contribution from (for example) the pole in the $\omega_1$ integral at $\omega_\alpha$ and the pole in the $\omega_2$ integral at $-\omega_\alpha$, the factor $(\omega_1+\omega_2)^{-1}$ in the expression (\ref{pfrule}) diverges. To resolve this problem we construct two different forms of the integrand. One will be used when calculating residues of pairs of poles that do not sum to zero. This expression is calculated as described above. When we calculate the residue of a pair of poles which add to zero, we use a different expression which is obtained without the partial fractioning step. We set $\omega_2=-\omega_1$ immediately, and then rewrite factors $\cos\varphi$ in the numerator using the second expression in Eq. (\ref{num-example}). 

The non trivial integrals that we need can be written
\bea
\label{3-ints-a}
&& I^{00}_{010}(\hat\omega_1,\hat\omega_2,\vec k)  
=\int^{2\pi}_0 d\varphi \; \frac{1}{D_-(\hat\omega_1)} \equiv  I(\hat\omega_1,\vec k)\,,
\\ \label{3-ints-b}
&& I^{00}_{020}(\hat\omega_1,\hat\omega_2,\vec k)  
=\int^{2\pi}_0 d\varphi \; \frac{1}{\big[D_-(\hat\omega_1)\big]^2}  \equiv J(\hat\omega_1,\vec k) \,,
\\ \label{3-ints-c}
&& I^{00}_{110}(\hat\omega_1,\hat\omega_2,\vec k)  =\int^{2\pi}_0 d\varphi \; \frac{C(\varphi)}{D_-(\hat\omega_1)}  \equiv K(\hat\omega_1,\vec k)\,.
\eea
We also need $I^{00}_{001}(\hat\omega_1,\hat\omega_2,\vec k) = I(\hat\omega_2,-\vec k)$ and $I^{00}_{101}(\hat\omega_1,\hat\omega_2,\vec k) = K(\hat\omega_2,-\vec k)$. The three integrals (\ref{3-ints-a}, \ref{3-ints-b}, \ref{3-ints-c}) can be calculated analytically. The results are 
\bea
&& I(\hat\omega,\vec k) =\frac{2 \pi }{\sqrt{\hat\omega  + i \epsilon -\sqrt{1-x^2} } 
\sqrt{\hat\omega + i \epsilon +\sqrt{1-x^2} }} \,, 
\\[2mm]
&& J(\hat\omega,\vec k) = \frac{2 \pi  \,\hat\omega}
{\left(\hat\omega  + i \epsilon -\sqrt{1-x^2} \right){}^{3/2} 
\left(\hat\omega  + i \epsilon +\sqrt{1-x^2}\right){}^{3/2}} \,,
\\[2mm]
&& K(\hat\omega,\vec k) = \frac{2\pi}{\hat\omega^2-1}\,\bigg[\frac{\hat\omega}{\sqrt{x^2+m_{\rm min}^2}} - \frac{1}{\sqrt{\hat\omega  + i \epsilon -\sqrt{1-x^2} } 
\sqrt{\hat\omega + i \epsilon +\sqrt{1-x^2} }}\bigg] \,.
\eea

\end{document}